\documentclass{aa}
\usepackage{color}
\usepackage{colortbl}
\usepackage{multirow}
\usepackage{dcolumn}
\usepackage{amsmath}
\usepackage{amssymb}
\usepackage{graphicx}
\usepackage{epsfig}
\usepackage{changebar}
\usepackage{natbib}
\usepackage{multirow}
\usepackage{subfigure}
\usepackage{txfonts}
\usepackage{relsize}

\usepackage{longtable,lscape}
\usepackage[figuresright]{rotating}
\newcolumntype{d}[1]{D{.}{\cdot}{#1}}

\newcolumntype{.}{D{.}{.}{-1}}

\newcommand{\msun}{M$_\odot$}
\newcommand{\lsun}{L$_\odot$}

\newcommand{\kms}{km~s$^{-1}$}
\newcommand{\Tr}{$T^*_{\rm{R}}$}

\newcommand{\Trv}{$\int T_{\rm{R}}^*~\rm{d}V$}

\newcommand{\cmthree}{cm$^{-3}$}

\newcommand{\vlsr}{$V_{\rm{LSR}}$}

\begin{document}
   \title{The RMS Survey:}

   \subtitle{$^{13}$CO observations of candidate massive YSOs in the northern Galactic plane}

   \author{J.~S.~Urquhart
          \inst{1}
          \and
			 A.~L.~Busfield
			 \inst{1}
			 \and
			 M.~G.~Hoare
			 \inst{1}
			 \and
			 S.~L.~Lumsden
			 \inst{1}
			 \and
			 R.~D.~Oudmaijer
			 \inst{1}
			 \and
			 T.~J.~T.~Moore
			 \inst{2}
			 \and
			 A.~G.~Gibb
			 \inst{3}
			 \and
			 C.~R.~Purcell
			 \inst{4,5}
			 \and
			 M.~G.~Burton
			 \inst{4}
			 \and
			 L.~J.~L.~Mar\'echal
			 \inst{6}
			 \and
			 Z.~Jiang
			 \inst{7}
			 \and
			 M.~Wang
			 \inst{7}
          }

   \offprints{J. S. Urquhart: jsu@ast.leeds.ac.uk}

   \institute{School of Physics and Astrophysics, University of Leeds, Leeds, LS2~9JT, UK 
         \and
             Astrophysics Research Institute, Liverpool John Moores University, Twelve Quays House, Egerton Wharf, Birkenhead, CH41~1LD, UK 
			\and
			Department of Physics and Astronomy, University of British Columbia, 6224 Agricultural Road, Vancouver, BC, V6T 1Z1, Canada 
			\and
				School of Physics, University of New South Wales, Sydney, NSW 2052, Australia  
			\and 
			Jodrell Bank Observatory, University of Manchester, Cheshire, SK11~9DL, UK
			\and
			\'Ecole Normale Sup\'erieure,
D\'epartement de Physique, 24 rue Lhomond, F-75005, Paris, France	
         \and
			Purple Mountain Observatory, Nanjing, P. R. China    
				 }

   \date{}

\abstract
   {The Red MSX Source (RMS) survey is an ongoing multi-wavelength observational programme designed to return a large, high-resolution mid-infrared colour-selected sample of massive young stellar objects (MYSOs). We have identified $\sim$2000 MYSO candidates located within our Galaxy by comparing the colours of MSX and 2MASS point sources to those of known MYSOs. The aim of our follow-up observations is to identify other objects with similar colours such as ultra compact (UC) HII regions, evolved stars and planetary nebulae (PNe) and distinguish between genuine MYSOs and nearby low-mass YSOs.}      
   {A critical part of our follow-up programme is to conduct $^{13}$CO molecular line observations in order to determine kinematic distances to all of our MYSO candidates. These distances will be used in combination with far-IR and (sub)millimetre fluxes to determine bolometric luminosities which will allow us to identify and remove nearby low-mass YSOs. In addition these molecular line observations will help in identifying evolved stars which are weak CO emitters.}
   {We have used the 15~m James Clerk Maxwell Telescope (JCMT), the 13.7~m telescope of the Purple Mountain Observatory (PMO), the 20~m Onsala telescope and the 22~m Mopra telescope to conduct molecular line observations towards 508 MYSOs candidates located in the 1st and 2nd Quadrants. These observations have been made at the $J$=1--0 (Mopra, Onsala and PMO) and $J$=2--1 (JCMT) rotational transition frequency of $^{13}$CO molecules and have a spatial resolution of $\sim$20\arcsec--55\arcsec, a sensitivity of $T_{\rm{A}}^*$ $\simeq$ 0.1~K and a velocity resolution of $\sim$0.2~\kms. We complement these targeted observations with $^{13}$CO spectra extracted from the Galactic Ring Survey (GRS), which have a velocity resolution of $\sim$ 0.21 \kms\ and sensitivity $T_{\rm{A}}^*$ $\simeq$ 0.13-0.2~K, towards a further 403 RMS sources.}
   {In this paper we present the results and analysis of the $^{13}$CO spectra obtained towards 911 MYSO candidates. We detect $^{13}$CO emission towards 780 RMS sources which corresponds to approximately 84\% of those observed. A total of 2595 emission components are detected above 3$\sigma$ level (typically $T^*_{\rm{A}} \ge 0.3$~K), with  multiple components being observed towards the majority of these sources -- 520 sources ($\sim$56\%) -- with an average of $\sim$4 molecular clouds detected along each line of sight. These multiple emission features make it difficult to assign a unique kinematic velocity to many of our sample. We have used archival CS ($J$=2--1) and maser velocities to resolve the component multiplicity towards 175 sources ($\sim$20\%) and have derived a criterion which is used to identify the  most likely component for a further 191 multiple component sources. Combined with the single component detections we have obtained unambiguous kinematic velocities for 638  of the 780 MYSOs candidates towards which CO is detected ($\sim$80\% of the detections). The 141 sources for which we have not been able to determine the kinematic velocity will require additional line data. Using the rotation curve of Brand and Blitz (1993) and their radial velocities we calculate kinematic distances for all detected components.}
	{}

\authorrunning{J. S. Urquhart et al.}
\titlerunning{$^{13}$CO observations of MYSO candidates}
\maketitle
\section{Introduction}
\subsection{Background}

The formation and evolution of massive stars ($>$10$^4$~\lsun) is still cloaked in mystery despite the fact that they are responsible for some of the most energetic astrophysical phenomena. Massive stars can have an enormous impact on their local environment through their UV radiation, molecular outflows and strong stellar winds as well as depositing huge amounts of processed material back to the ISM. Their impact is not limited to their local environment but can also influence the evolution of their whole galaxy. They are also thought to play an important role in regulating star formation through the propagation of strong shocks into their natal molecular cloud and nearby clouds. These shocks may be responsible for triggering subsequent generations of star formation or disrupting conditions necessary for star formation in nearby clouds (\citealt{elmegreen1998}).

Our understanding of how massive stars form and evolve lags behind that of low-mass stars, which in broad terms is relatively well understood (\citealt{shu1987,shu1993}). There are several reasons for this; firstly, massive stars are extremely rare and have very short evolutionary time scales. This results in them generally being located  much farther away than sites of low-mass star formation. Secondly, massive stars are known to form exclusively in clusters making it difficult to identify and attribute derived quantities to individual sources. Finally, they reach the main sequence whilst still accreting material, and are heavily embedded within their natal molecular cloud, hidden away beneath many magnitudes of visual extinction. However, given their profound impact, not only on their local environment, but also on a galactic scale, it is vital that we understand the environmental conditions and processes involved in their birth and the early stages of their evolution.

Due to the observational difficulties, until relatively recently, the only catalogue of well characterised massive young stellar objects (MYSOs) had been limited to 30 or so serendipitously detected sources (\citealt{Henning1984}) which are mostly nearby. The situation has improved considerably in recent years with a number of studies using a variety of selection criteria (e.g., \citealt{molinari1996,walsh1997,sridharan2002}). These studies have identified many new MYSOs, however, they are all based on IRAS colours which bias them towards bright, isolated sources and tend to avoid dense clustered environments and the Galactic mid-plane where the majority of MYSOs are expected to be found -- the scale height of massive stars is $\sim$30\arcmin\  (\citealt{reed2000}). The limited number of MYSOs identified by these studies, and the way they have been selected, means that they are probably not representative of the general MYSO population. This makes statistical studies of many aspects of massive star formation difficult.

\subsection{The RMS Survey}

There is an obvious need for a large, high-resolution mid-infrared colour-selected sample with sufficient numbers of sources in each luminosity bin to allow statistical studies as a function of luminosity. The Red MSX Source (RMS) survey is an ongoing observational programme designed to return just such a sample with complementary data with which to address these issues.

The formation of massive stars, and the earliest stages of their evolution,
take place inside cold ($\sim$15~K), dense (H$_2$ $>$10$^4$~cm$^{-3}$),
compact ($<$0.5~pc) clumps located within giant molecular clouds.
Observationally these early stages should be characterised by strong
emission at (sub)millimetre wavelengths due to cold dust and gas, with
little or no emission at mid-infrared wavelengths as they have not yet begun
to warm their surrounding cocoon of dust (\citealt{beuther2007}).  (Sub)millimetre continuum observations of infrared dark clouds (IRDCs;
\citealt{rathborne2006}), towards IRAS selected sources (e.g.,
\citealt{faundez2004,beltran2006}, and towards complete molecular complexes (e.g., \citealt{motte2007}), have detected a significant
number of cold, dense clumps and infrared quite cores.   These cold cores
have similar masses ($\sim$100~\msun) and densities ($>$10$^5$~cm$^{-3}$) as
warmer cores found towards known high-mass star forming regions, which has led to speculation they may represent the some of the earliest stages of massive
star formation yet observed.

As the growing protostar begins to warm up its external environment it
enters what is known as the hot molecular core stage as exotic
molecules are evaporated from the surrounding dust grains
(\citealt{cesaroni2005}). Since the Kelvin-Helmholtz time scale is
much shorter than the free-fall time, massive stars can begin fusing
hydrogen in their cores while still deeply embedded within their
infalling envelope.  As the envelope is dispersed the object emerges
as a bright point source in the mid- and then near-infrared. This is the
MYSO stage where the infrared luminosity has reached 10$^4$--10$^5$~L$_\odot$,
consistent with young O and early B-type stars (\citealt{wynn-williams1982}). However, they do not yet ionise their surroundings, probably due to the ongoing accretion (see discussion by \citealt{hoare2007b,hosokawa2008}). They are
usually associated with massive bipolar molecular outflows (e.g.,
\citealt{wu2004}) ionised stellar winds that are weak thermal radio
sources ($\sim$1~mJy at 1 kpc; \citealt{hoare2002}). Finally, when the
Lyman continuum radiation does escape the object enters the UCHII region
phase, which expands and clears the remaining molecular cloud material
in the immediate vicinity (\citealt{hoare2007a}).

\citet{lumsden2002} compared the colours of sources from the MSX and 2MASS point source catalogues to those of known MYSOs to develop a colour selection criteria from which we identified approximately 2000 MYSO candidates spread throughout the Galaxy ($|b|<$5\degr). Sources toward the Galactic centre were excluded (defined as $|l|<10$\degr) due to confusion and difficulties in calculating kinematic distances. 

A problem with a colour selected sample is that the shape of the spectral energy distribution from an optically thick cloud is insensitive to the type of heating source. Therefore there are several other types of embedded, or dust enshrouded object, that have similar colours to MYSOs and contaminate our sample, such as ultra compact (UC) HII regions, evolved stars and a small number of planetary nebulae (PNe). 

The RMS survey is a multi-wavelength programme of follow-up observations designed to distinguish between genuine MYSOs and these other embedded or dusty objects (\citealt{hoare2005,urquhart2007c}) and to compile a database of complementary multi-wavelength data with which to study their properties.\footnote{http://www.ast.leeds.ac.uk/RMS.} These include
high resolution cm continuum observations to identify UCHII regions and PNe (\citealt{urquhart2007a}), mid-infrared imaging to identify genuine point sources, obtain accurate astrometry and avoid excluding MYSOs located near UCHII regions (e.g., \citealt{mottram2007}), and near-infrared spectroscopy (e.g., \citealt{clarke2006}) to distinguish between MYSOs and evolved stars. 

Another aspect of our follow-up campaign is to obtain kinematic distances which are crucial for calculating luminosity and hence distinguishing between nearby low- to intermediate-mass YSOs and genuine MYSOs. We have conducted a programme of $^{13}$CO observations towards all RMS sources not previously observed or for which good quality data was not available (see Sect.~\ref{sect:grs} for details). In addition to being able to determine distances towards all sources where line profiles are detected, these observations allow us to eliminate a significant number of evolved stars ($\sim$10--15\%) since they are not generally associated with strong CO emission ($^{12}$CO $J$=1--0 and $J$=2--1 typically less than 1~K; \citealt{loup1993}).
In a recent paper (\citealt{urquhart2007b}; hereafter Paper I) we presented the results of $^{13}$CO observations made towards 854 RMS source located in the southern Galactic plane (i.e., 180\degr$<l<$350\degr). In this paper we present the spectral line data towards 911 RMS sources located in the northern Galactic plane (NGP;10\degr$<l<$180\degr). These spectra have been obtained from a combination of 508 targeted observations and 403 spectra extracted from the Galactic Ring Survey (\citealt{jackson2006}).\footnote{http://www.bu.edu/galacticring/.} In Sect.~2 we outline our source selection, observational and data reduction procedures. We present our results and statistical analysis in Sect.~\ref{sect:results}. In Sect.~\ref{sect:summary} we present a summary of the results and highlight our main findings. 

\section{Observational strategy}
\label{sect:strategy}

The $^{13}$CO transition was chosen for these targeted observations as it is generally found to be only moderately optically thick and is thus a better tracer of column density than $^{12}$CO. Moreover, using $^{13}$CO avoids many of the problems often encountered when observing with $^{12}$CO, which is nearly always optically thick and can often result in complex structures with multiple components and/or heavily self-absorbed spectral profiles, particularly towards the inner Galaxy (c.f. \citealt{jackson2006,wu2001}). The CS transition was considered as its high critical density (excitation threshold $\sim$10$^4$--$10^5$~\cmthree) makes it an excellent tracer of high density molecular gas. However, CS is much less abundant than $^{13}$CO and therefore requires significantly longer on-source integration time to obtain an equivalent signal to noise. 

To avoid re-observing sources that had previously been observed we began by conducting a literature search from which we were able to identify approximately a hundred sources for which good data was already available (see Sect.~\ref{sect:literature} for details). The next step was to try to take advantage of the many large scale surveys which are concentrating on the northern GLIMPSE region (GLIMPSE is one of the Spitzer Legacy projects; see \citealt{benjamin2003} for details). One such survey is the Galactic Ring Survey (GRS; see \citealt{jackson2006} for details) which has recently been completed. This project mapped a large region of the northern GLIMPSE field in $^{13}$CO ($J$=1--0) and is therefore directly applicable to our project. We extracted spectra towards approximately four hundred RMS sources which lie within the GRS survey region (see Sect:~\ref{sect:grs} for details). Finally, targeted observations were made towards the remaining RMS sources that are located outside the GRS region and for which no archival data was available. 

\subsection{Literature search}
\label{sect:literature}

Of the 2000 or so MYSO candidates identified from our selection criteria (hereafter the RMS catalogue) 1021 are located within the 1st and 2nd Galactic Quadrants. Only about a quarter of these have unconfused IRAS PSC fluxes and therefore most of our sources are currently unknown in the literature. To reduce the number of targeted observations needed we searched the literature using SIMBAD\footnote{http://simbad.u-strasbg.fr/simbad/sim-fid.} to identify sources for which good quality molecular line data were already available. 

In total we identified 324 sources towards which molecular line observations have previously been reported. The majority of the molecular line data found towards RMS sources was drawn from two large observational programmes: 1) a CS ($J$=2--1) survey towards an IRAS colour-selected sample (proposed by \citealt{wood1989}) of suspected UCHII regions reported by \citet{bronfman1996}; and 2) a CO survey by \citet{wouterloot1989} which concentrated mainly on IRAS sources located in the outer Galaxy. 

A total of 253 sources were found to be in common with sources observed by \citet{bronfman1996}, 49 of which they report as non-detections. The majority of their detections were excluded from our observations, however, because CS is generally harder to detect than $^{13}$CO we re-observed all of their non-detections. In addition to their non-detections we re-observed $\sim$25\% of their detections to allow for consistency checks and comparison of the two tracers. 

\citet{wouterloot1989} used the $^{12}$CO~($J$=1--0) transition to observe 1302 IRAS selected sources located within 10\degr\ of the Galactic plane and between 85\degr\ $< l <$ 280\degr. We find 110 sources in common between their observations and our northern MYSO candidates. Some of these were excluded but a number were found to possess complex profiles, or were found to be saturated, which led to a large number being re-observed as part of our programme.

\subsection{Archival data: The Galactic Ring Survey}
\label{sect:grs}

The GRS survey was conducted by Boston University using the Five College Radio Astronomy Observatory (FCRAO), 14~m telescope located in New Salem, Massachusetts. The observations were made at the $^{13}$CO ($J$=1--0) frequency between December 1998 and March 2005 using the single sideband focal plane array receiver SEQUOIA (\citealt{erickson1999}). At the $^{13}$CO frequency the FCRAO has a FWHM beam size of 46\arcsec.

The GRS survey covers a total area of 75.4 deg$^2$ between 18\degr $<l<$55.7\degr\ and $|b|<$1\degr. For the observations of the $l>40\degr$ region a bandwidth of 40~MHz with 512 channels was used, which corresponds to a velocity resolution of 0.21~\kms\ and total velocity coverage of $\sim$100~\kms. These observations were centred at a \vlsr\ of 40~\kms\ providing coverage between $-$10 to 90~\kms. For the region $l<40\degr$ a dual channel correlator was used allowing two independent intermediate frequencies to be observed simultaneously each with a bandwidth of 50~MHz and 1024 channels. These were centred at 20 and 100~\kms, overlapping by 270 channels and providing a total velocity bandwidth of 210~\kms\ with a velocity resolution of 0.13~\kms. Therefore the  processed data cubes available have slightly different ranges of \vlsr, for those with Galactic longitudes $l<40\degr$ the \vlsr\ range = $-$5 to 135~\kms\ and for those with Galactic longitudes $l>40\degr$ the \vlsr\ range = $-$5 to 85~\kms. The velocity range covered by the GRS is outlined in green on the lower panel of Fig.~\ref{fig:rms_distribution}. 

In total 403 RMS sources were found to be located within region covered by the GRS survey. Spectra for all of these sources were extracted from the GRS data cubes. The data cubes are fully sampled with a pixel size of 22\arcsec, however, to increase the signal to noise we averaged the emission over nine pixels (3$\times$3 region) centred on the position of each RMS source effectively reducing the angular resolution to $\sim$1\arcmin but giving a similar sensitivity to the targeted observations ($\sim$0.1~K; see next section).

\subsection{Targeted observations}
\label{sect:observations}

\begin{sidewaysfigure*}
\begin{center}

\includegraphics[height=0.95\textwidth,angle=270, trim=0 3 0 0]{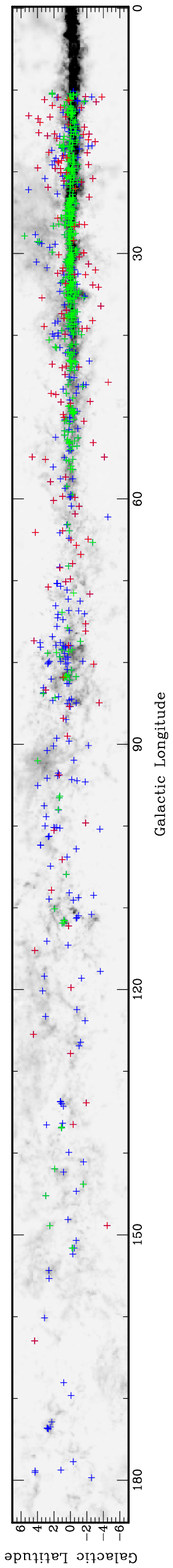}\\
\includegraphics[height=0.95\textwidth, angle=270, trim=0 0 0 3]{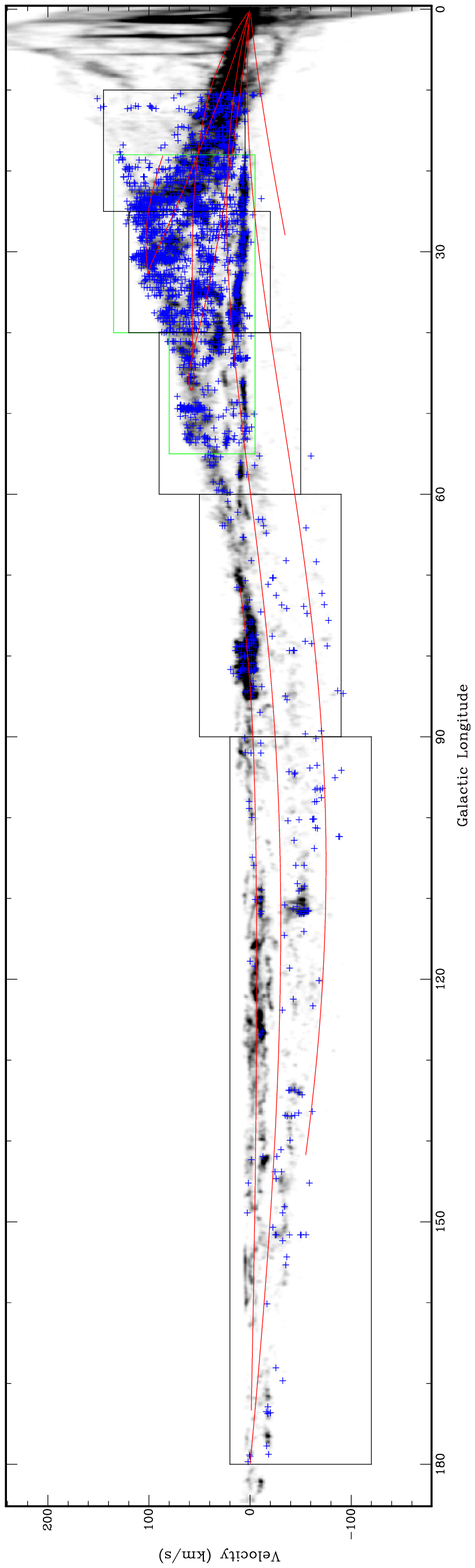}

\caption{Top panel: Galactic distribution of all northern hemisphere RMS sources located in the 1st and 2nd quadrants observed in $^{13}$CO and presented in this paper. The colours correspond to whether an observation resulted in a non-detection (red), a single component was detected (blue) or multiple components are detected (green). The integrated $^{12}$CO emission map of \citet{dame2001} (grey scale) shows the Galactic distribution of molecular material. Bottom panel: Galactic longitude-velocity plot showing the velocities of all detected $^{13}$CO components as a function of Galactic longitude. The distribution of molecular material is shown in grey scale (again from \citealt{dame2001}) for comparison. The location of the spiral arms taken from model by \citet{tayor1993} and updated by \citet{cordes2004} are over-plotted in red. The area outlined in green indicates the longitude and velocity range covered by the GRS (see text for details). The regions covered by our targeted observations made with the Mopra telescope are outlined in black.}

\label{fig:rms_distribution}
\end{center}
\end{sidewaysfigure*}

We made $^{13}$CO observations towards 508 RMS sources from a total of 1021 located in the northern Galactic plane. These sources were observed using four different telescopes, namely: the 15~m James Clerk Maxwell Telescope (JCMT) located on Mauna Kea, Hawaii\footnote{The James Clerk Maxwell Telescope is operated by The Joint Astronomy Centre on behalf of the Science and Technology Facilities Council (STFC) of the United Kingdom, the Netherlands Organisation for Scientific Research, and the National Research Council of Canada.}; 13.7~m telescope of the Purple Mountain Observatory (PMO) located at the Qinghai station; 20~m Onsala telescope located approximately 50~km south of Goteborg\footnote{Onsala is operated by the Swedish National Facility for Radio Astronomy, Onsala Space Observatory at Chalmers University of Technology.} and the 22 m Mopra telescope located near Coonabarabran, New South Wales, Australia\footnote{Mopra is operated by the Australia Telescope National Facility, CSIRO and the University of New South Wales during the time of the observations.}. Just over half of these sources were observed in $^{13}$CO ($J$=1--0) (i.e., all sources observed with either the PMO, Onsala or Mopra telescopes), however, the instruments available on the JCMT are unable to operate at this frequency and therefore the higher excitation $^{13}$CO ($J$=2--1) transition was used. The use of these two different transitions should not affect our findings since YSOs are expected to have similar intensities at these frequencies (\citealt{little1994}).

All of these observations were performed in position-switching mode, with typical on-source integration times of $\sim$10~minutes for the PMO, Onsala and Mopra observations, and $\sim$6~minutes for the JCMT observations (these integration times were reduced for stronger sources). The on-source integration time was split into a number of separate scans consisting of 1 minute of on- and off-source integrations. Reference positions were offset from source positions by 1 degree in a direction perpendicular to the Galactic plane. These were chosen to avoid contamination by emission in the reference position at a similar velocity. In some cases, particularly towards the Galactic centre, several positions needed to be tried before a suitable reference position was found. A summary of the individual telescopes' observational parameters and number of sources observed is presented in  Table~\ref{tbl:CO_radio_parameters}. 

The PMO has the ability to observe $^{12}$CO, $^{13}$CO and C$^{18}$O ($J$=1--0) transition simultaneously. However, in this paper we only present the $^{13}$CO data for the following reasons: first, the velocity resolution of the $^{12}$CO is three times worse than that of the other two lines and this, combined with the higher optical depth, makes it much harder to distinguish between clouds at similar velocities. Second, due to the lower abundance of the C$^{18}$O isotopomer and relatively short integration time used for these observations, the signal-to-noise ratios in these data were much lower than in the $^{13}$CO data and often resulted in non-detections. The velocity range covered by the $^{12}$CO data is three times that of the other two lines ($\sim$350~\kms) which proved useful during the observations for identifying the \vlsr\ of emission regions along the line of sight and allowed the $^{13}$CO to be centred more effectively.

\begin{table*}[!tbp]
\begin{center}
\caption{Observational parameters for the $^{13}$CO observations.}
\label{tbl:CO_radio_parameters}
\begin{minipage}{\linewidth}
\begin{tabular}{lcccccc}
\hline
\hline

Project 	&\multicolumn{4}{c}{RMS}& \multicolumn{2}{c}{GRS} \\
\hline
Telescope& Mopra & Onsala & PMO &JCMT & \multicolumn{2}{c}{FCRAO}\\
Galactic Longitude range (deg)&10--52&54--180 & 10--90&10--180 & $l>40$& $l<40$\\

\# of sources observed & 62 & 83 & 134&229& \multicolumn{2}{c}{403}\\
Rest frequency (GHz) & 110.201& 110.201& 110.201& 220.3986 &\multicolumn{2}{c}{110.201}\\
Total bandwidth (MHz)& 64 & 80 & 43 & 267 & 40 &2$\times$50 \\
Vel. bandpass (\kms) & 174 & 224 & 117&360 &100&210\\ 
Number of channels   & 1024 & 1600 &1024& 1713 & 512 &2$\times$1024\\
Vel. resolution (km s$^{-1}$) & 0.17 & 0.14 & 0.11& 0.21 & \multicolumn{2}{c}{0.21}\\
Beam size (\arcsec) & 33 & 35 & 55&21 &\multicolumn{2}{c}{46}\\
Date of observations & 2002-2005 & March 2003&Jan 2005 &2003-2004&\multicolumn{2}{c}{Dec 1998-Mar 2005}\\
Integration time (mins) & 10 & 10 & 10& 6 &\\
Typical Tsys  (K)          & 310 & 420 & 280 & 510 &\multicolumn{2}{c}{310} \\     
Telescope efficiency ($\eta_{\rm{MB}}$) & 0.55 & 0.43 &0.61& 0.69 & \multicolumn{2}{c}{0.48}\\
\hline
\end{tabular}\\
\\
\end{minipage}
\end{center}
\end{table*}

Typical system temperatures obtained for both the PMO and Mopra observations are between 230--330~K depending on weather conditions and telescope elevation, with values typically below 300~K. The observations made from Onsala had system temperatures approximately 100~K higher. The higher frequency observations conducted with the JCMT have system temperatures between 450--550~K. Telescope pointing was regularly checked ($\sim$2--3~hours) and found to be better than 3\arcsec\ for the JCMT and 10\arcsec\ for the PMO, Onsala and Mopra telescopes. To correct the measured antenna temperatures for atmospheric absorption, ohmic losses and rearward spillover and scattering (i.e., to place the data on the corrected antenna temperature scale $T^*_{\rm{A}}$), a measurement was made of an ambient load (assumed to be at 290~K) following the method of \citet{kutner1981}; this was done approximately every two hours.

In the upper panel of Fig.~\ref{fig:rms_distribution} we present a plot of the distribution of molecular material as a function of Galactic longitude and latitude as traced by the \citet{dame2001} $^{12}$CO survey. The coloured crosses indicate the RMS positions observed as part of the $^{13}$CO observations presented here; the crosses are coloured red, blue and green to indicate whether the observation resulted in a non-detection, detection of a single or multiple components respectively. The RMS sources towards which $^{13}$CO emission is detected correlate extremely well with the distribution of the $^{12}$CO. In the lower panel of Fig.~\ref{fig:rms_distribution} we present a plot of the distribution of $^{12}$CO as a function of Galactic longitude and V$_{\rm{LSR}}$ (again taken from \citealt{dame2001}). We have over-plotted the velocities of all detected components. Additionally, we have overlaid the longitude-velocity of the updated spiral arm model of \citet{tayor1993} by \citet{cordes2004} in red to illustrate how their velocities vary as a function of longitude in our Galaxy.

The velocities over which molecular clouds can be found in the northern Galactic plane range from approximately $-100$~\kms\ to 150~\kms. Since the total Galactic velocity range ($\sim$250~\kms) is larger than the total velocity bandpass available to either the Mopra or Onsala observations it was necessary to change the central velocity as a function of Galactic longitude in order to optimise our velocity coverage. We used the Galactic longitude-velocity distribution maps of \citet{dame2001} to centre the available velocity bandpass as a function of longitude to cover as much of the northern Galaxy plane as possible. The central velocities used at various Galactic longitudes  and the region covered by each are outlined in black on the lower panel of Fig.~\ref{fig:rms_distribution}. Comparing the regions covered by our observations with that of the distribution of $^{12}$CO shows that, although our velocity bandpass is limited, we have obtained an almost complete coverage of the northern Galactic plane.


\subsection{Data reduction}

\begin{figure}[!]
\begin{center}
\includegraphics[width=0.9\linewidth]{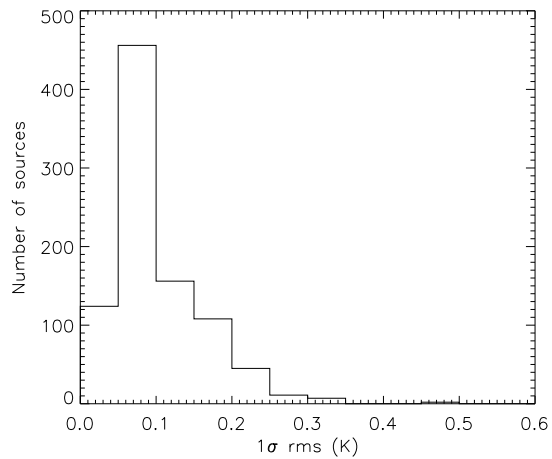}

\caption{Histograms of 1$\sigma$ rms obtained from 6--10 minute observations using a bin size of 0.05 K. Over 90\% of the observations have a sensitivity of better than 0.15~K with typical values of $\sim$0.1~K.}

\label{fig:rms_hist}
\end{center}
\end{figure}

For the targeted observations the basic spectral line data reduction and processing was performed using standard procedures and the software package specific to the individual telescopes were used: \textsc{DFM} (Data
From Mopra)\footnote{A tcl/tk graphical interface written for
\textsc{SPC} (Spectral Line Reduction Package) by C. Purcell.} for
Mopra data, the \textsc{XS}\footnote{Written by P. Bergman, Onsala
Space Observatory.} package for Onsala data and
\textsc{CLASS}
(Continuum and Line Analysis Single-dish Software)\footnote{Part of the \textsc{GILDAS} (Grenoble
Image and Line Data Analysis Software) working group software.} for the JCMT
and PMO data. As previously mentioned the observation of each source results in between six and ten separate integrations. Sky emission was subtracted from these individual scans which were inspected to remove poor data. The remaining scans were averaged together to produce a single spectrum for each source and a low-order polynomial was subtracted from the baseline. The resulting 1$\sigma$ sensitivity for each spectrum was typically $T^*_{\rm{A}} \simeq$ 0.1~K per channel (see Fig.~\ref{fig:rms_hist}).

Finally all of the data (GRS and targeted spectra) were scales to the main beam brightness temperature (\Tr), by dividing the antenna temperatures by the telescope efficiency ($\eta_{\rm{MB}}$; see Table~\ref{tbl:CO_radio_parameters} for individual telescope efficiencies).
Since the main objective of these observations is to obtain kinematic velocities, rather than accurate line intensities, little absolute calibration was performed. We therefore estimate the calibration uncertainties could be as much as a factor of two.

\section{Results and analysis}
\label{sect:results}

\subsection{Detection statistics}

From the combination of data extracted from the GRS survey and our targeted observations we have obtained spectra towards 911 RMS sources. Of these we detect $^{13}$CO emission towards 780 RMS sources which corresponds to $\sim$84\% -- this compares well with the 88\% detection rate found for the southern sample (see Paper I for details). Approximately 56\% of all detections have resulted in two or more components, with 2595 components being recorded above 3$\sigma$ (typically $T^*_{\rm{A}} \ge 0.3$~K).

\begin{figure*}[!t]
\begin{center}

\includegraphics[width=0.44\linewidth]{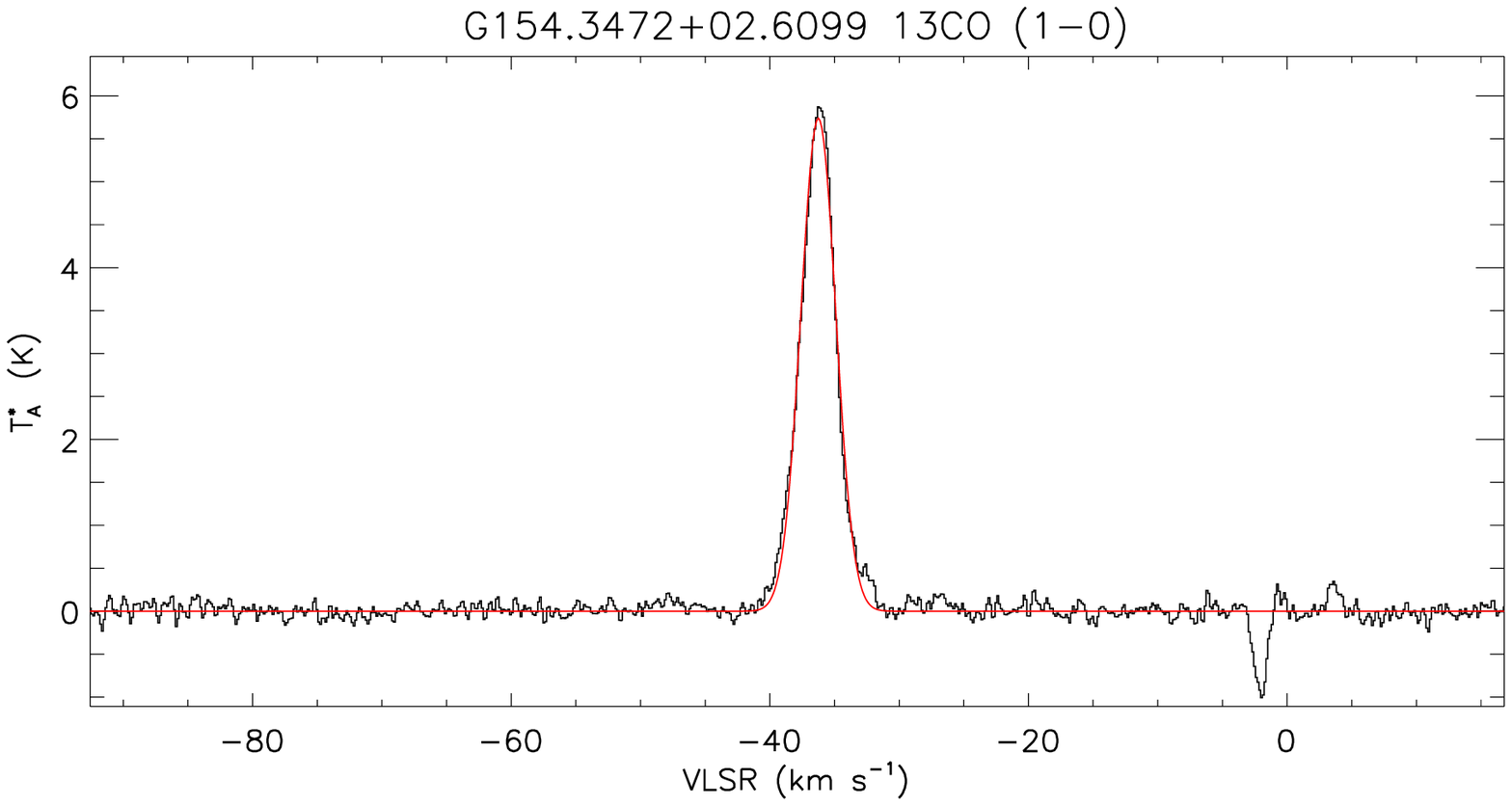} 
\includegraphics[width=0.44\linewidth]{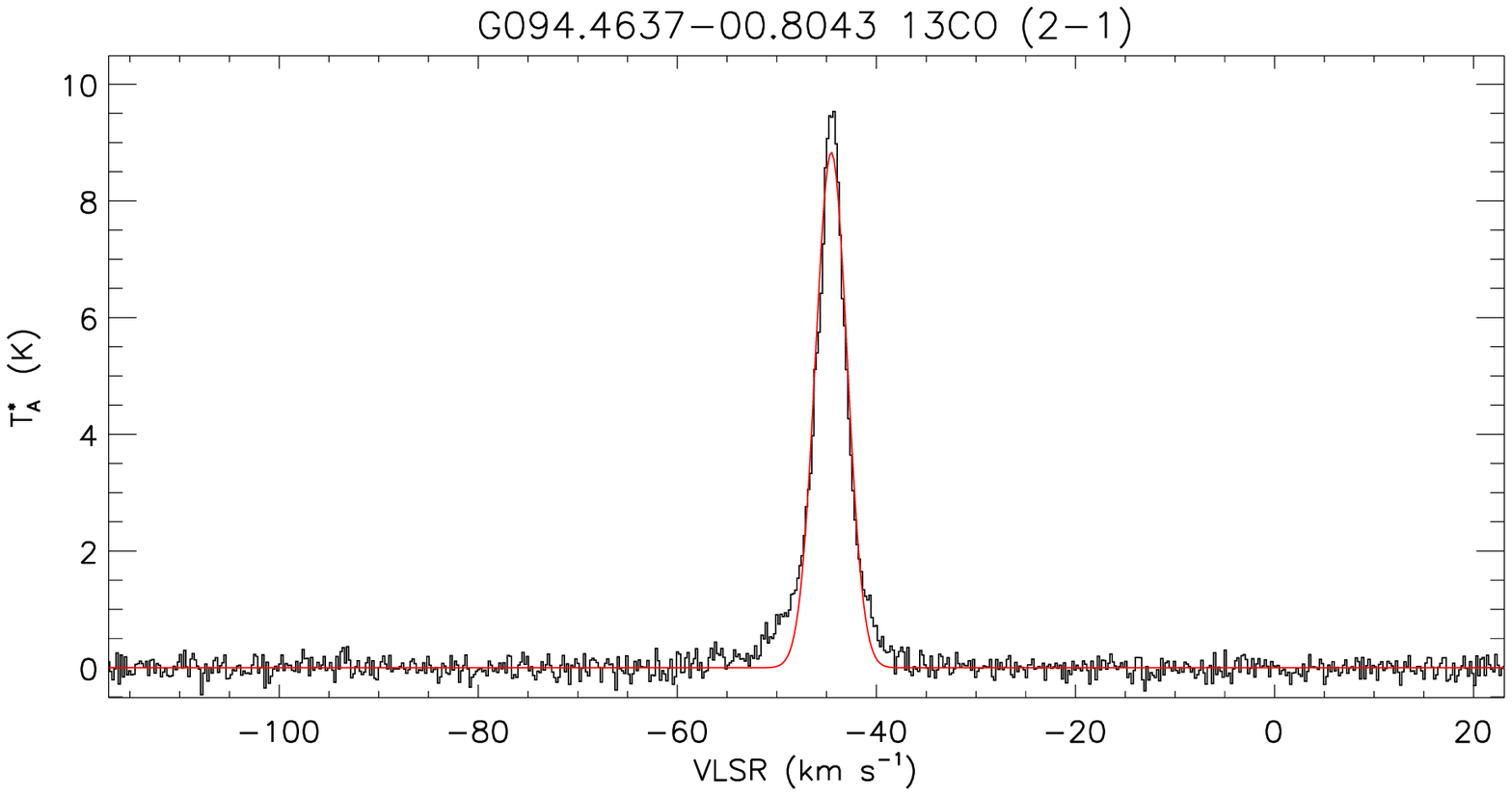} 
\includegraphics[width=0.44\linewidth]{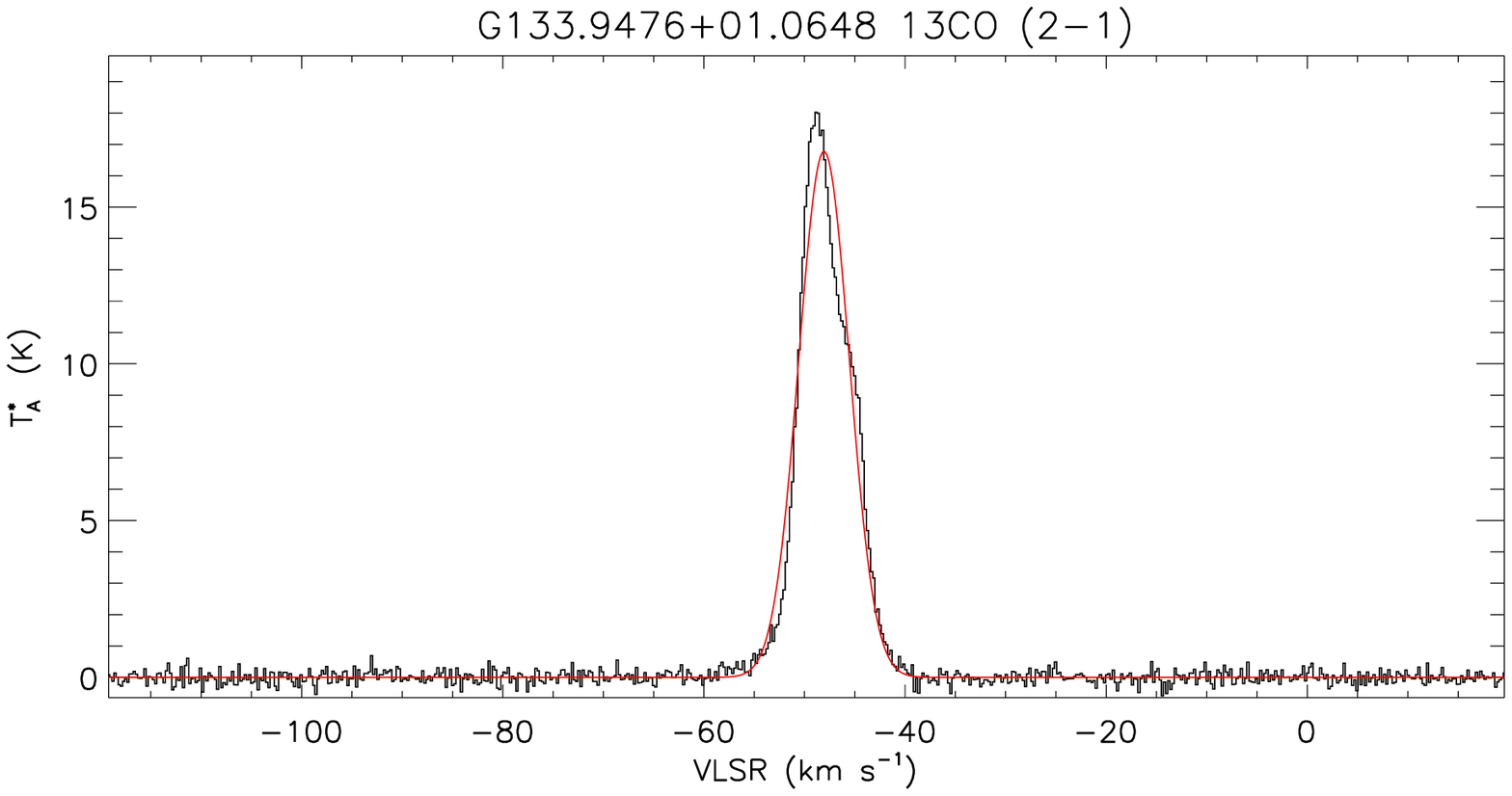} 
\includegraphics[width=0.44\linewidth]{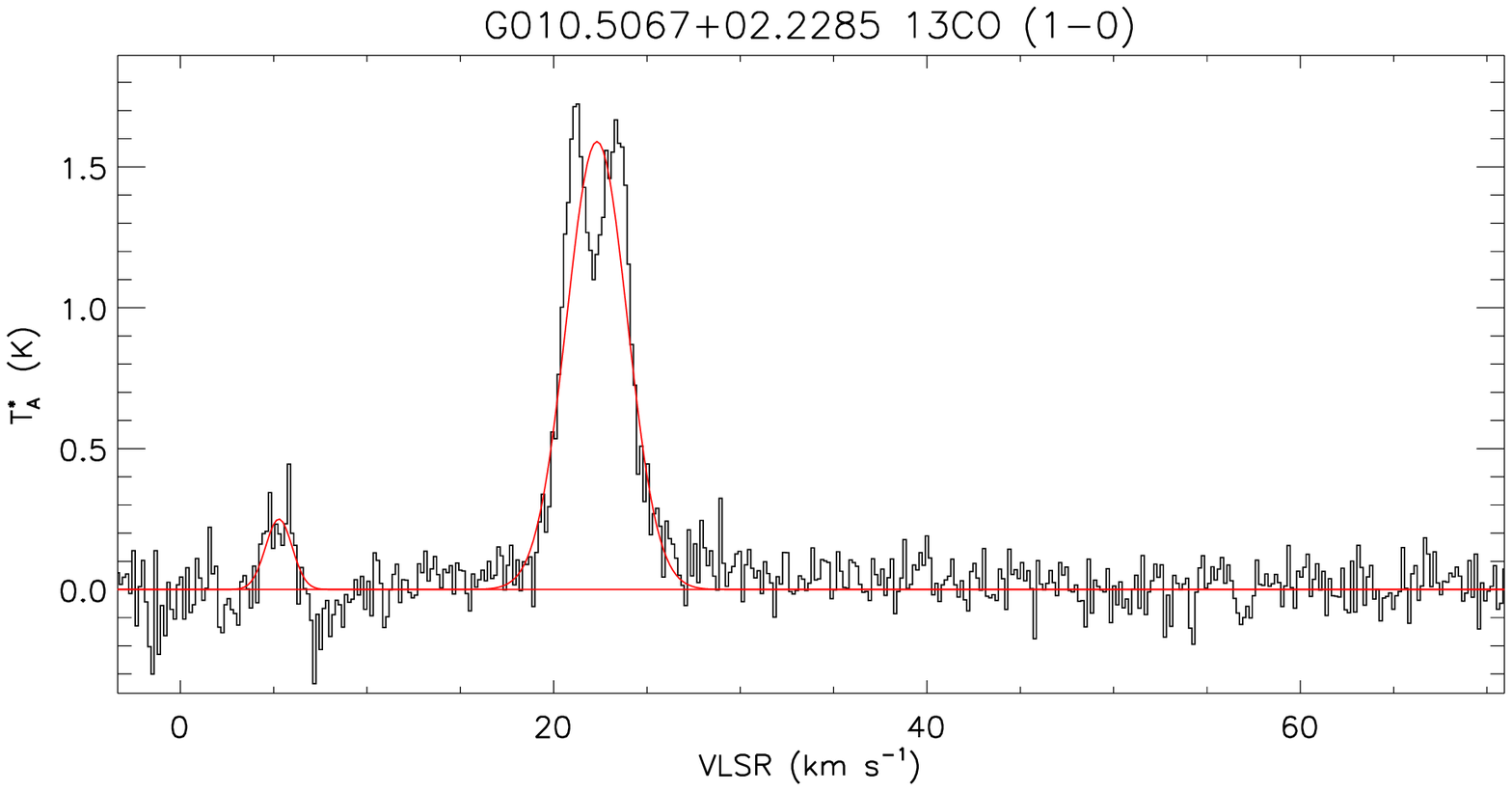} 
\includegraphics[width=0.44\linewidth]{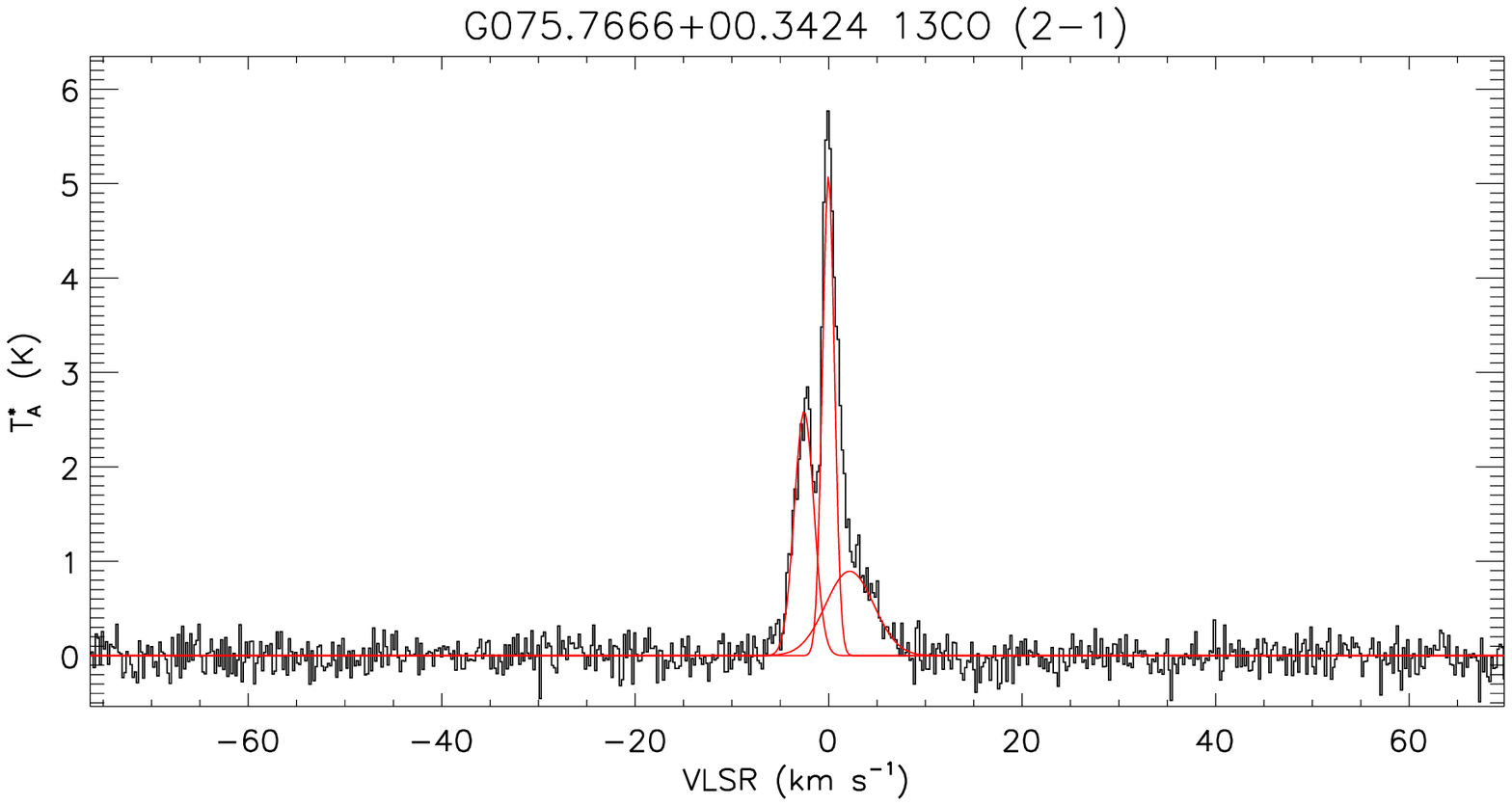} 
\includegraphics[width=0.44\linewidth]{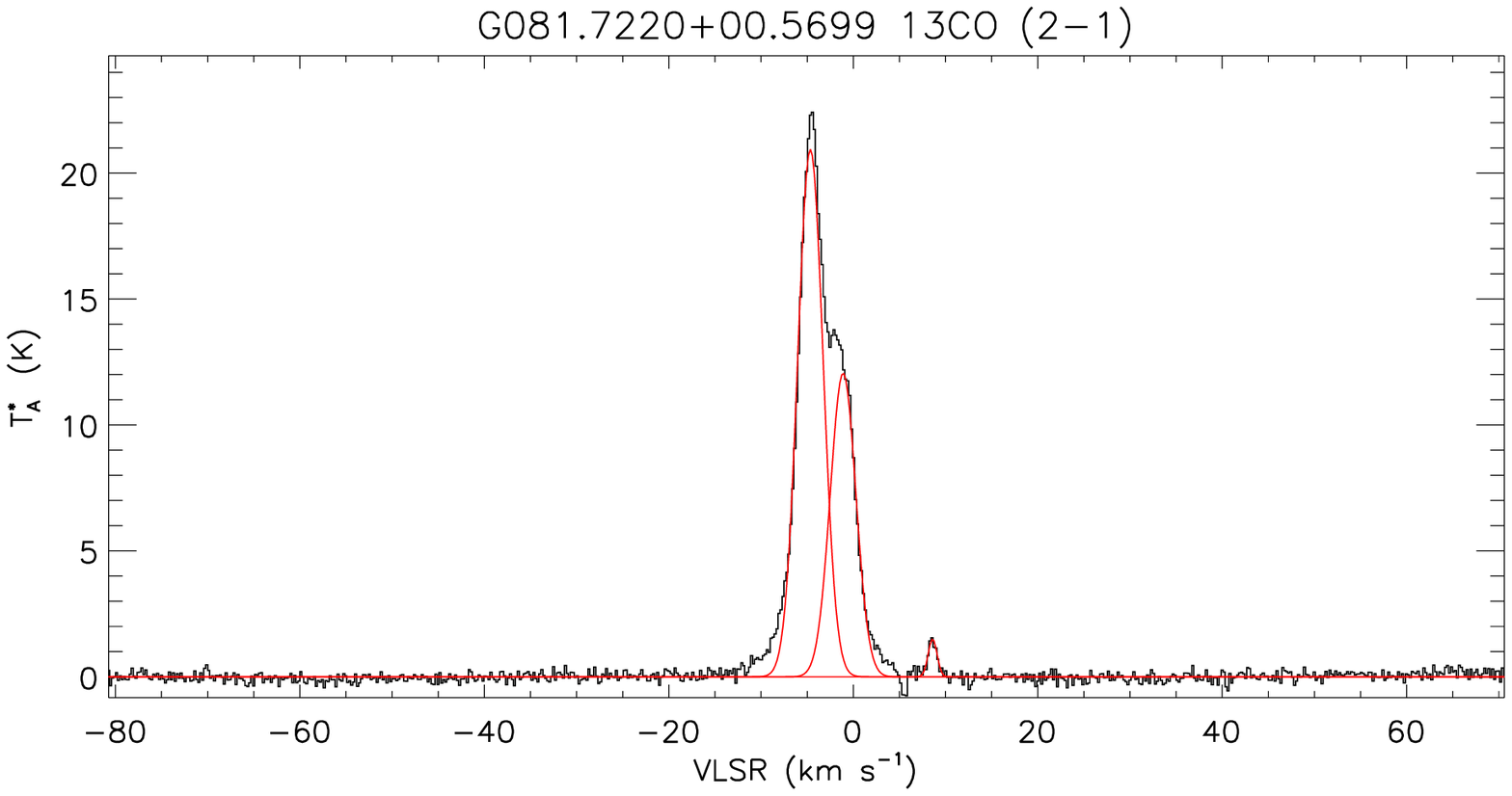} 

\includegraphics[width=0.44\linewidth]{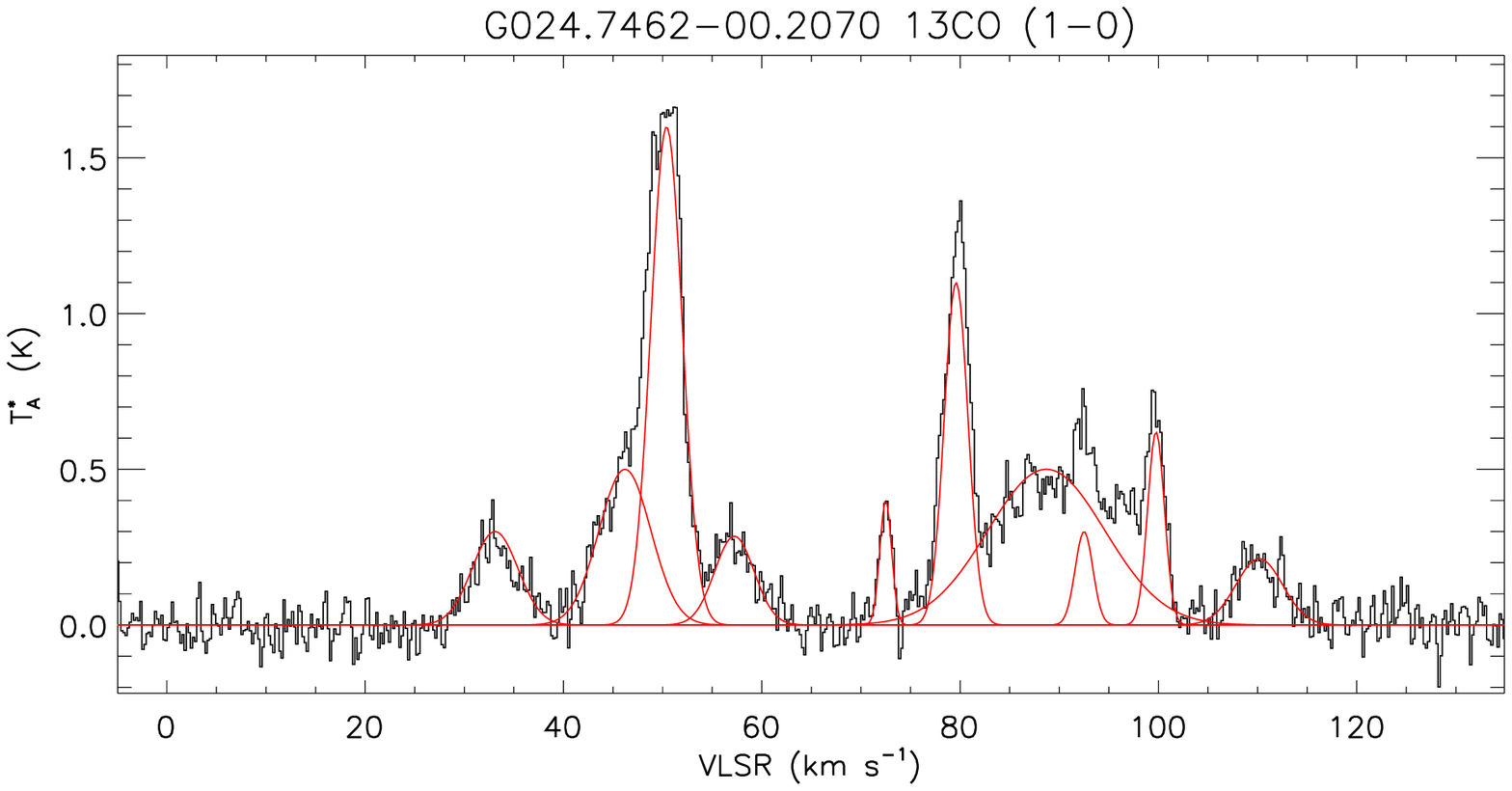}
\includegraphics[width=0.44\linewidth]{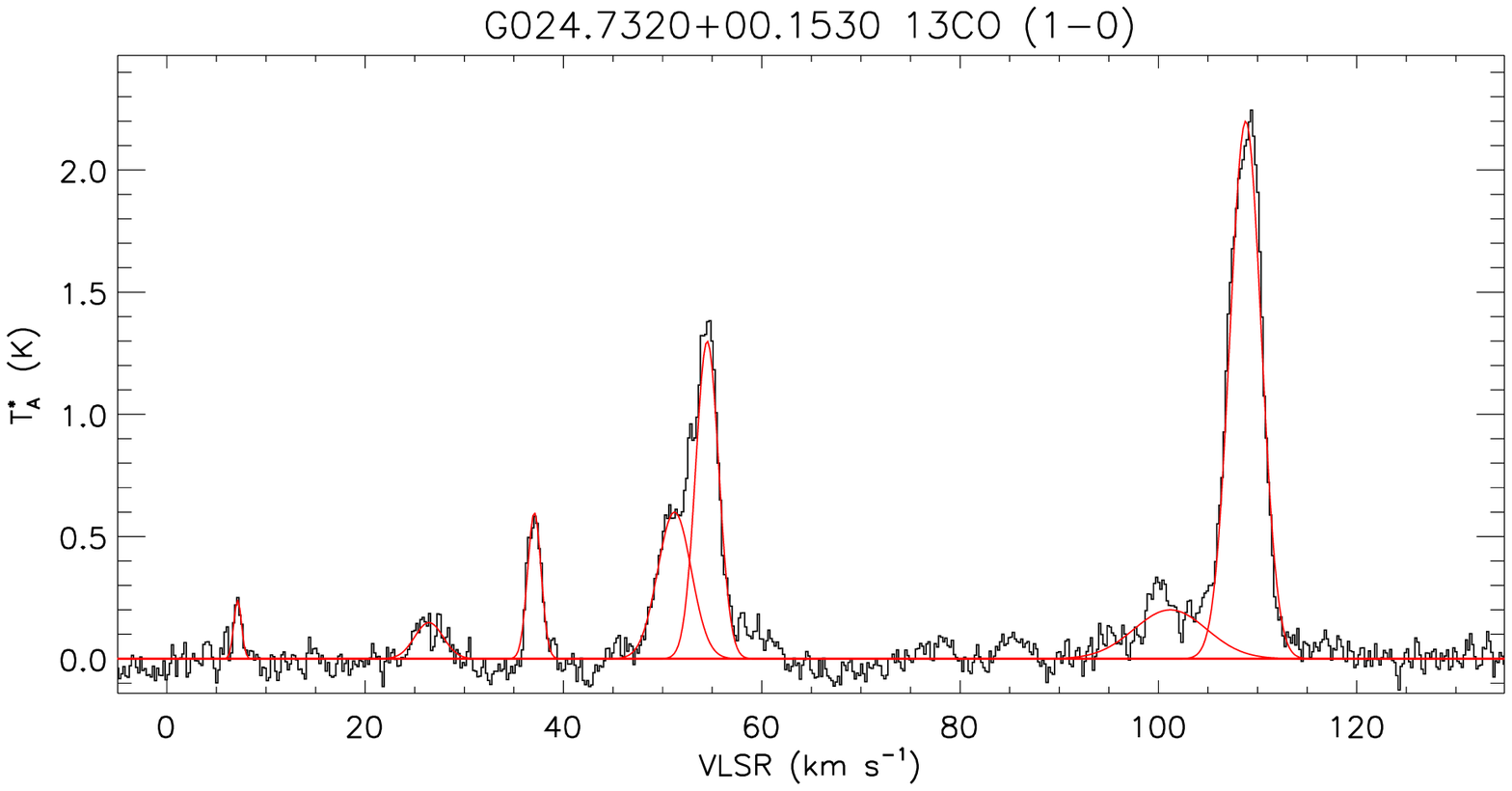}
\includegraphics[width=0.44\linewidth]{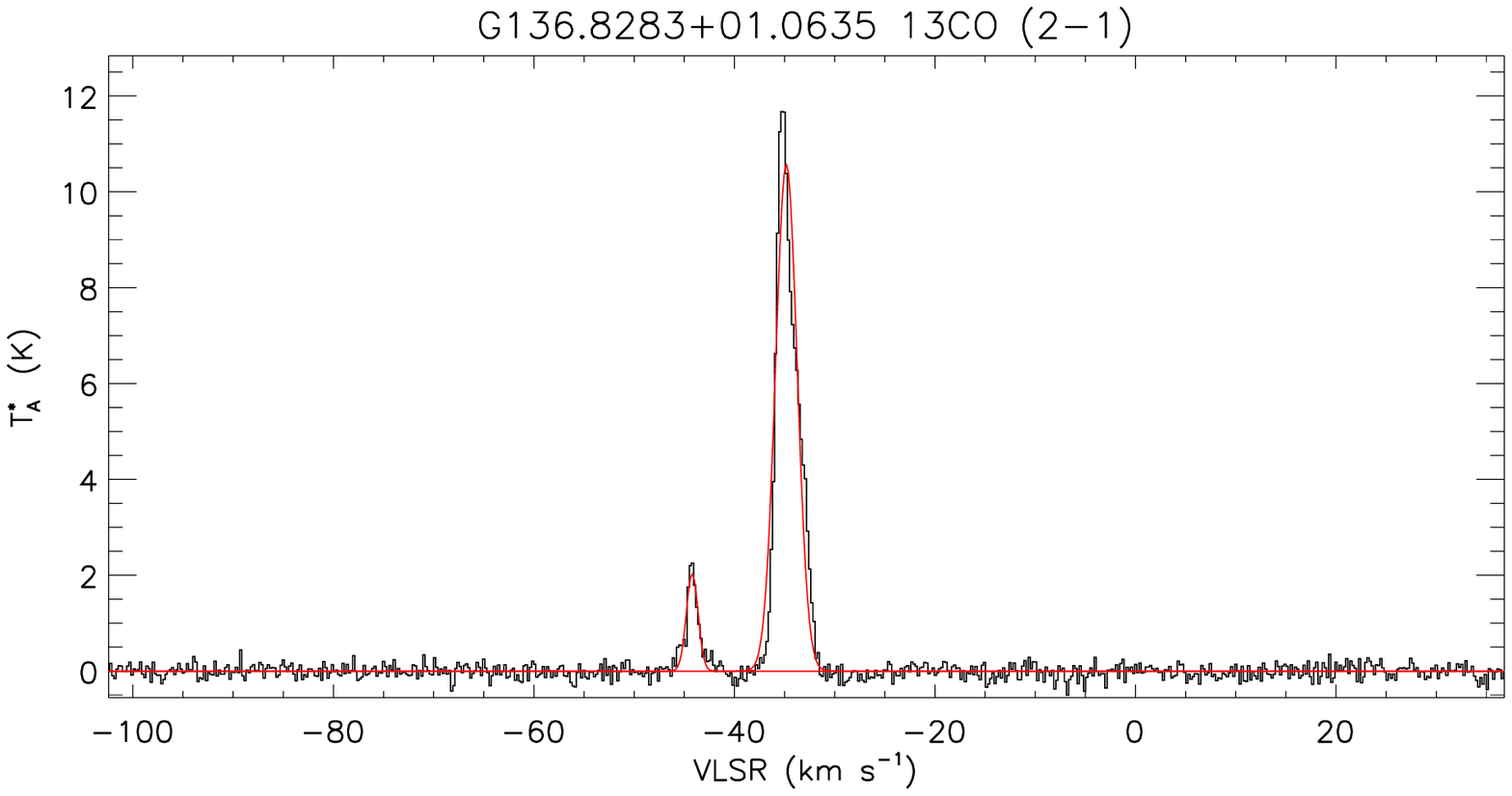}
\includegraphics[width=0.44\linewidth]{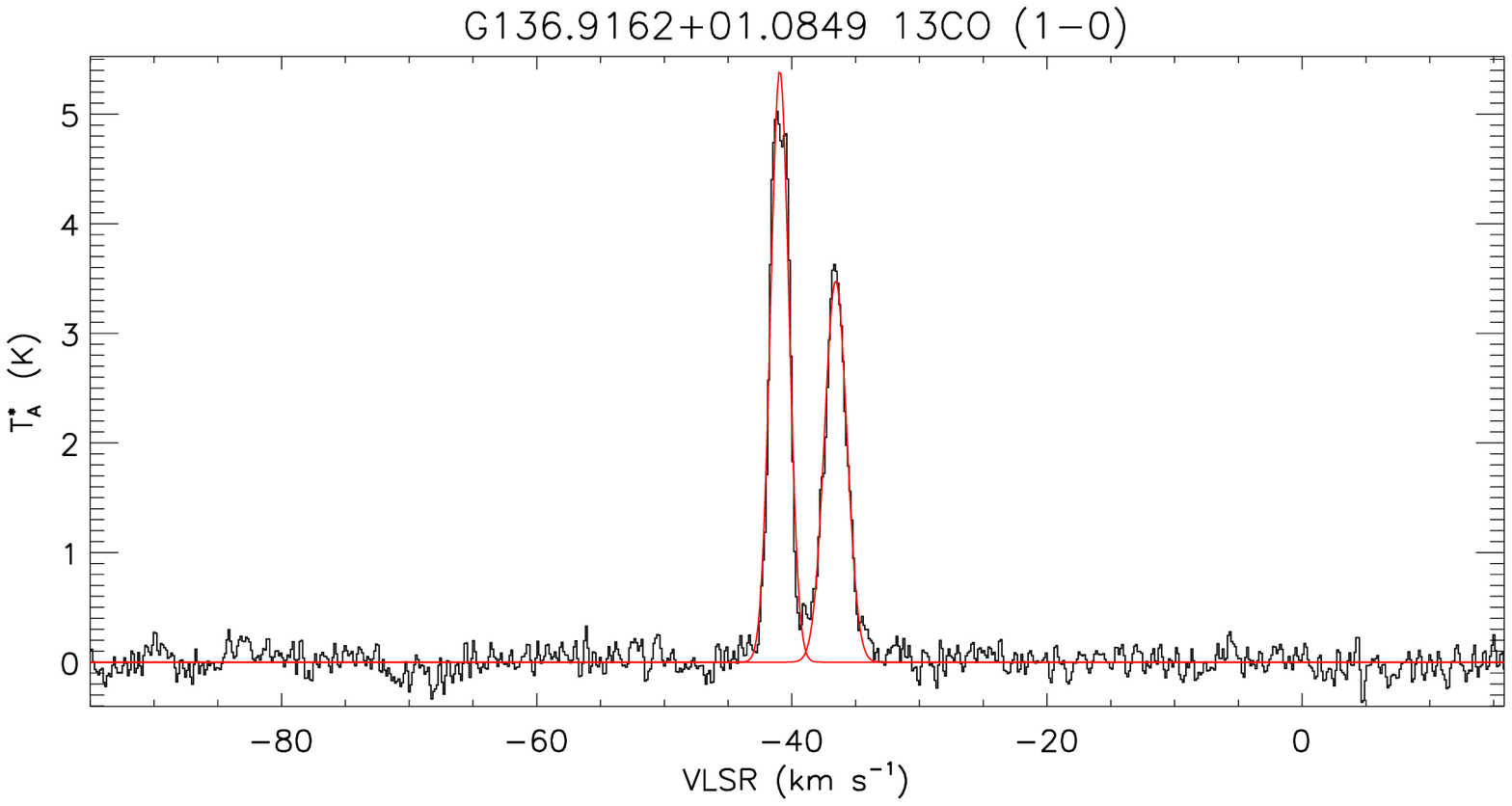}

\caption{Example spectra observed towards RMS sources. The spectra presented in the top three rows have been chosen to illustrate the different kinds of emission profiles exhibited in the data and to clarify the classification scheme used for the non-Gaussian profiles. Moving left to right from the top left the profiles are classified as follows: Gaussian, blue wing, red shoulder, self-absorbed, blended and blended feature with a main and shoulder component (see Sect. 3.1 for details). The bottom two rows show examples of multiple components seen towards the majority of our sample (see Sect. 3.3 for details).}

\label{fig:example_spectra}
\end{center}
\end{figure*}

The spectral line parameters were obtained by fitting Gaussian profiles to each component present in a given spectrum using \textsc{XS}. Where necessary, higher-order polynomials were fitted and subtracted from the baselines before the Gaussians were fitted. In the majority of cases a Gaussian profile provided a good fit to the data. However, in a significant number of cases, line components displayed significant deviation from a Gaussian profile to warrant an additional comment. We have adopted the classification scheme used by  \citet{wouterloot1989} for their outer Galaxy $^{12}$CO observations of colour selected IRAS  sources. They distinguish six different non-Gaussian profiles which are:  flat top, asymmetric blue or red top, blue or red asymmetry, self-absorption, blue and/or red wing, blue or red shoulder. 
In a number of cases we found two or more separate profiles overlapping with each other forming more complex profiles or where the shoulder component was particularly pronounced and a single Gaussian produced a poor fit to the data. In these cases it is not clear whether these represent the superposition of a number of clouds along the line of sight or supersonic motions within a particular cloud. Therefore in addition to the profiles just mentioned we identify a profile as blended if either fitted component overlaps with the other above the half-power level. In Fig.~\ref{fig:example_spectra} we present a sample of our spectra to illustrate the different line profile types and our fits to the data.

We present the parameters obtained from the Gaussian fits to the data and derived values in Table~\ref{tbl:source_parameters} and present plots of the fitted spectra in Fig.~\ref{fig:fitted_spectra}.\footnote{Full versions of Table~\ref{tbl:source_parameters} and Fig.~\ref{fig:fitted_spectra} are available in the online version.} The table format is as follows: the RMS names, positions and observations 1$\sigma$ rms in columns 1--4; component number and parameters extracted from Gaussian fits  (i.e., \vlsr, \Tr\ and FWHM) and profile type in columns 5--9; component integrated intensity in column 10; in columns 11--13 we present the radius from the Galactic centre (RGC) and kinematic distances determined from the \citet{brand1993} rotation model (see Sect.~\ref{sect:kinematic_distance} for details).

\begin{figure}[!t]
\begin{center}

\includegraphics[width=0.9\linewidth]{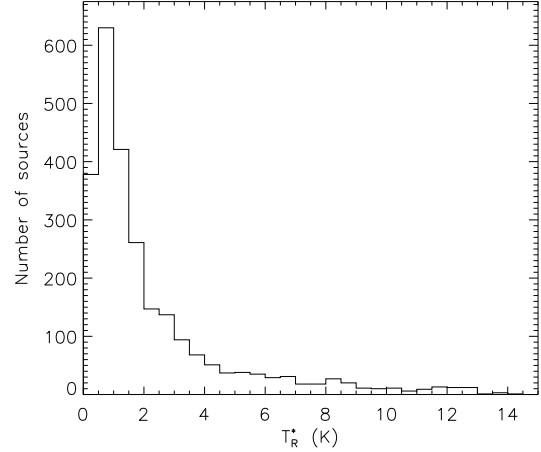}

\caption{Histogram of the distribution of \Tr\ measured towards RMS sources.  The bin size is 0.5~K.}

\label{fig:amp_hist}
\end{center}
\end{figure}

\begin{figure}[!t]
\begin{center}

\includegraphics[width=0.9\linewidth]{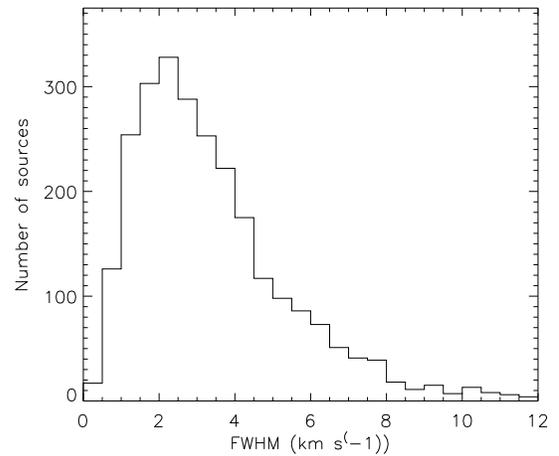}

\caption{Histogram showing the distribution of FWHM observed towards RMS sources. The distribution peaks at $\sim$2.5~\kms. Bin size used is 0.5~\kms.}

\label{fig:fwhm_hist}
\end{center}
\end{figure}
\begin{figure}[!t]
\begin{center}

\includegraphics[width=0.9\linewidth]{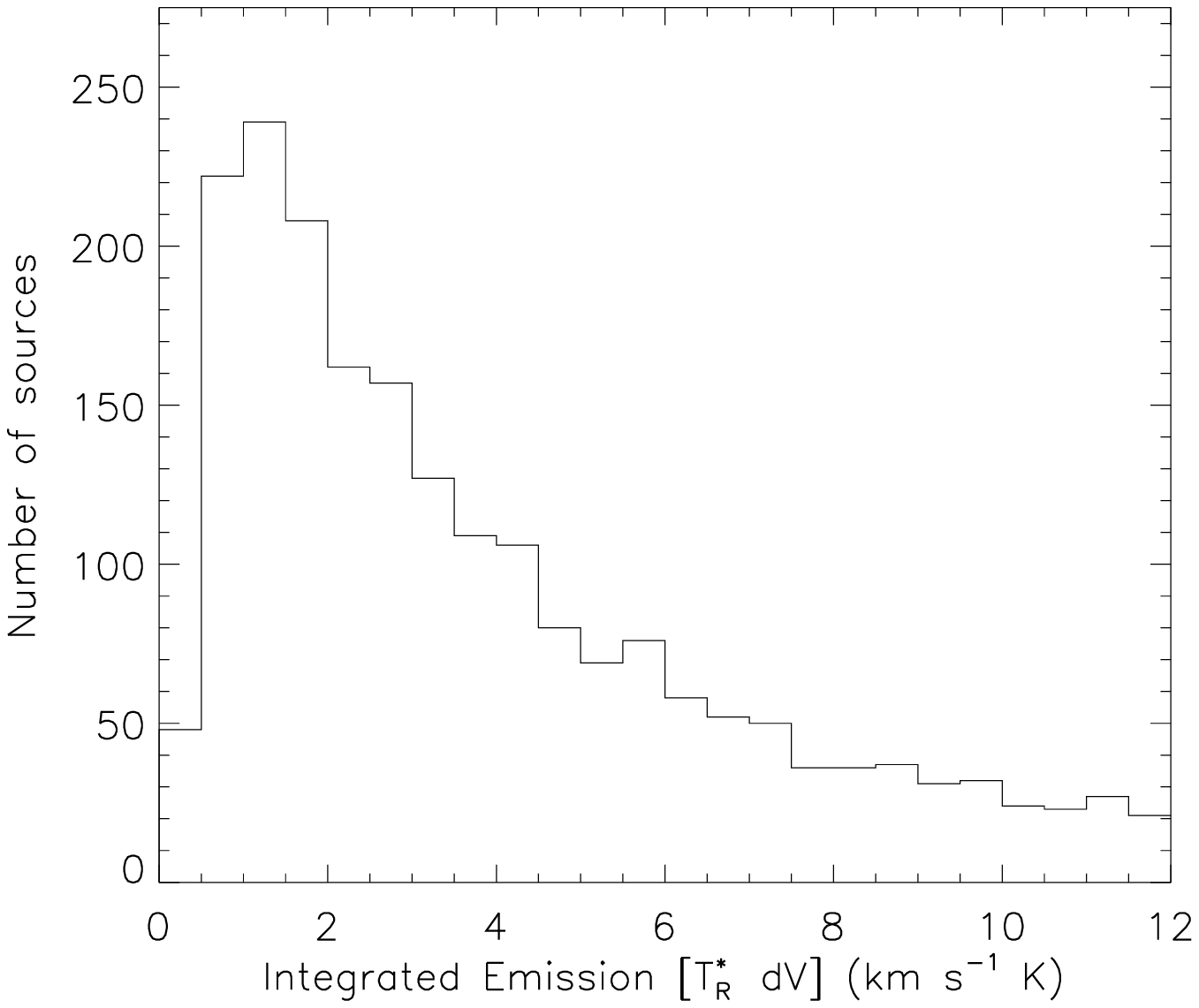}

\caption{Histogram of integrated intensity (\Trv) of all components detected towards RMS sources.}

\label{fig:int_data}
\end{center}
\end{figure}

In  Figs.~\ref{fig:amp_hist} and \ref{fig:fwhm_hist} we present histograms of the peak temperatures (\Tr) and FWHM of all detected components respectively.  The average antenna temperature is $\sim$1.5 K and FWHM is $\sim$3.5 \kms, however, both distributions are highly skewed to large values and consequently more typical (i.e., median) values  are $\sim$0.5 K and 2.5 \kms\ respectively. In Fig.~\ref{fig:int_data} we present the distribution of integrated intensities (\Trv = 1.06~\Tr~$\Delta$v, where $\Delta$v and \Tr\ are the line FWHM and peak intensity as determined from the fitted Gaussian components) of all detected components. 

\subsection{Galactic distribution}

\begin{figure*}
\begin{center}
\includegraphics[width=0.33\linewidth, trim=20 0 20 0]{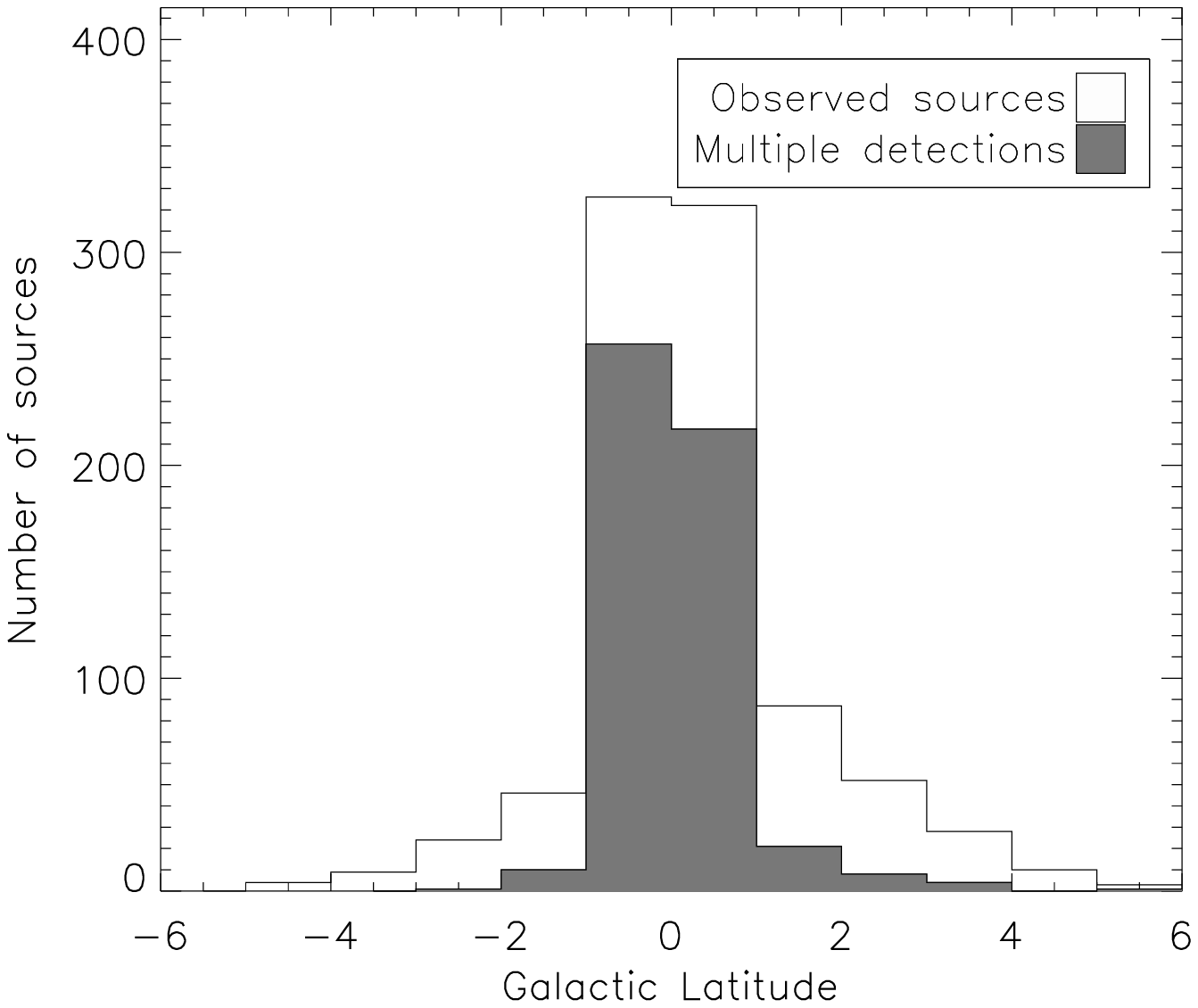}
\includegraphics[width=0.33\linewidth, trim=20 0 20 0]{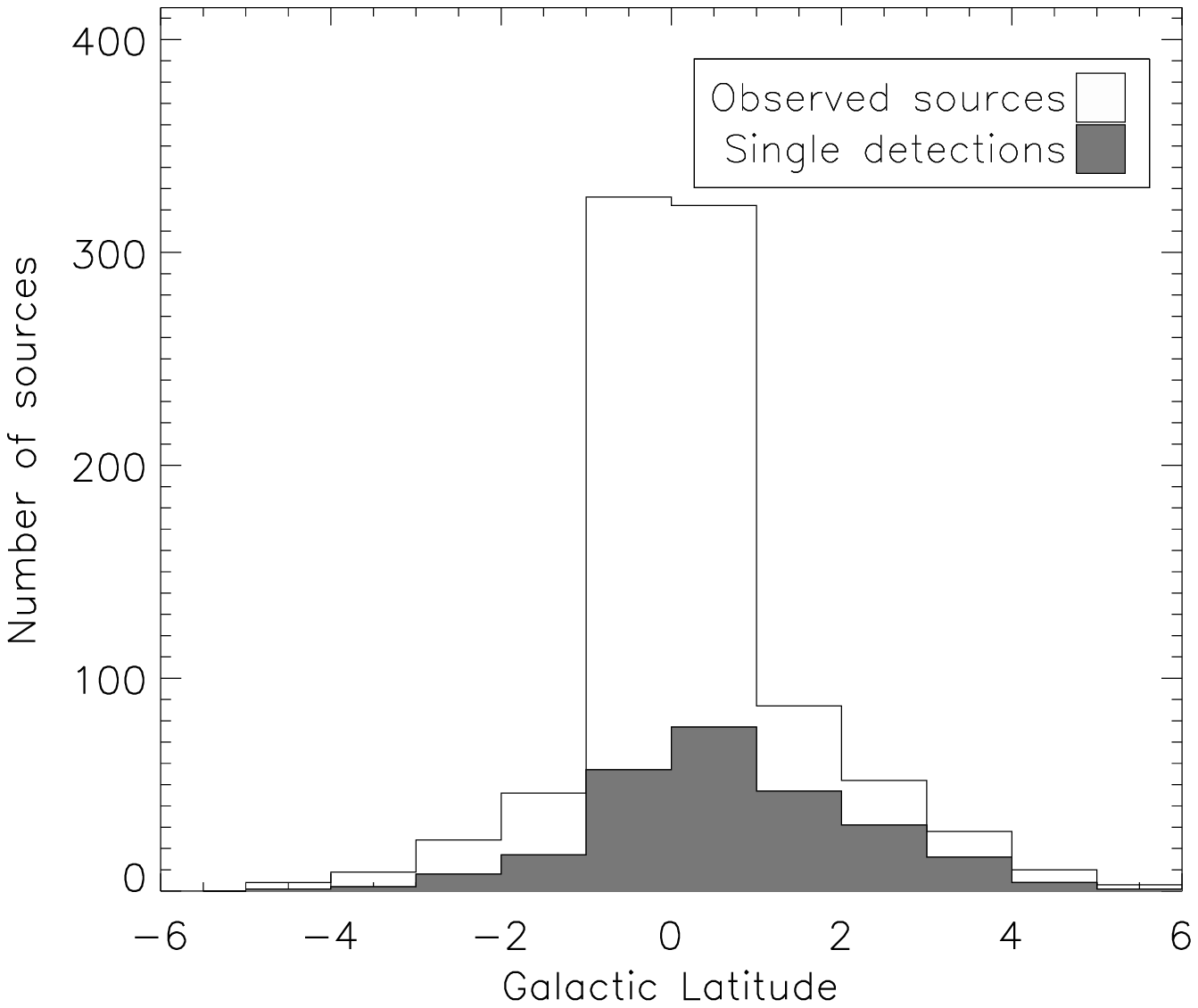}
\includegraphics[width=0.33\linewidth, trim=20 0 20 0]{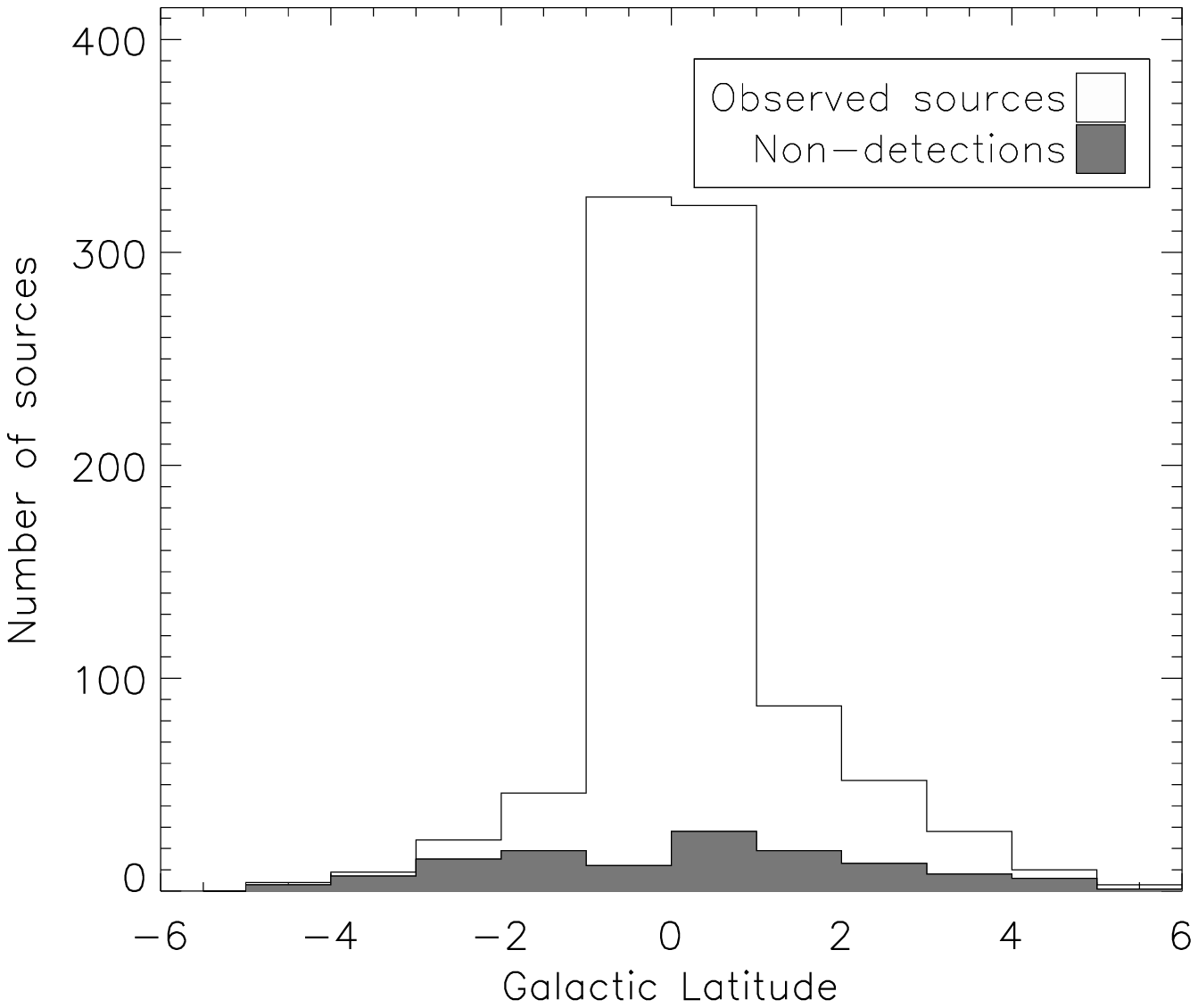}

\caption{Histograms of the Galactic latitude distribution of observed RMS sources. The unfilled histogram in each plot consists of all sources observed and reported in this paper. Histograms of the latitude distributions  of the multiple component detections, single component detections and non-detections are over-plotted (grey histogram) on the left, middle and right panels respectively. A bin size of 1 degree has been used for all of these histograms.}

\label{fig:lat_dist_histo}
\end{center}
\end{figure*}

The spatial distributions of the non-detections, single and multiple component detections can be seen in the Galactic longitude-latitude plot presented in the upper panel of Fig.~\ref{fig:rms_distribution}. The multiple component detections are concentrated towards the inner Galaxy and have a narrow latitude distribution. The single peak detections have a more even longitude distribution and are slightly wider latitude distribution compared to that of the multiple component detections. The distribution of the non-detections appears more evenly spread in both longitude and latitude, as would be expected if these sources represent some of the more evolved stars that contaminate our sample. The different latitude distributions of the three detection types is better illustrated in the histograms presented in Fig.~\ref{fig:lat_dist_histo}.
In these plots the overall distribution of the RMS sources presented in this paper is represented as an unfilled histogram, with the distributions of the multiple and single components detections and non-detections shown as filled histograms. The overall distribution of RMS sources covers a latitude range of $\sim$$\pm$6\degr\ with the vast majority fairly evenly distributed within $\pm$1\degr\ of the Galactic plane. The distribution of the multiple component detections is pretty much confined within 1\degr\ of the plane with a small tail out to $\sim$2\degr\ of the plane. The single component distribution is peaked between 0\degr\ and 1\degr\ and extends out to $\sim$5\degr\ above and below the plane. 

The non-detections appear to be fairly evenly distributed for latitudes greater than $|b|$ $>$ 1\degr, however, the distribution is weakly peaked between 0\degr\ and 1\degr\ and shows a minimum between $-1$\degr\ and 0\degr. These sources are most likely evolved stars which have similar colours to embedded YSOs but have very little, if any, associated molecular material. If these non-detections are indeed evolved stars we would expect them to be fairly evenly distributed in latitude.  Therefore the local minimum seen just below the plane could indicate a line of sight alignment between a small number of evolved stars and molecular clouds towards the Galactic mid-plane. However, these will be identified through our near infrared spectroscopic observations.

\begin{sidewaystable*}
\begin{minipage}[t][180mm]{\textwidth}
\caption{Sample table of source parameters.}
\label{tbl:source_parameters}
\centering
\begin{tabular}{@{}lcccl....l.ccc@{}} 
\hline\hline             
\multirow{2}{24mm}{Field Name$^a$} 	&	RA  		&Dec  		& rms & Id.$^b$	&	\multicolumn{1}{c}{\vlsr} 	& \multicolumn{1}{c}{\Tr}	& \multicolumn{1}{c}{FWHM} 	& \multicolumn{1}{c}{\Trv}&Profile type & \multicolumn{1}{c}{RGC} &Near	&Far & Notes \\ 
&	(J2000) 	& (J2000) 	& (K) & 		& \multicolumn{1}{c}{\kms} &\multicolumn{1}{c}{(T)} 	&\multicolumn{1}{c}{(\kms)} 	&  \multicolumn{1}{c}{(\kms\ K)}&				&\multicolumn{1}{c}{(kpc)}& \multicolumn{1}{c}{(kpc)}& \multicolumn{1}{c}{(kpc)} &  \\  
\hline

G012.7408+00.3305$^P$ & 18:12:08.6 & $-$17:43:51 & 0.12 & 1  $^\star$  &18.3 & 5.5 & 2.7 &   15.7 &  & 6.2 & 
2.4 (1.03)  & 14.2 (1.03) & 1 \\ 
 &  &  &  &2 &30.4 & 1.6 & 3.3 &    5.6 &  & 5.2 & 
3.4 (0.75)  & 13.1 (0.75) &  \\ 
G012.7465$-$00.1590$^P$ & 18:13:57.6 & $-$17:57:38 & 0.13 &1 &35.2 & 6.9 & 5.0 &   36.6 &  & 4.9 & 
3.8 (0.67)  & 12.8 (0.67) &  \\ 
 &  &  &  &2 &48.6 & 3.1 & 12.4 &   40.7 &  & 4.2 & 
4.5 (0.51)  & 12.0 (0.51) &  \\ 
G012.7499+00.3827$^P$ & 18:11:58.2 & $-$17:41:52 & 0.08 & 1  $^\star$  &18.1 & 6.4 & 3.1 &   21.0 & Red shoulder & 6.2 & 
2.4 (1.04)  & 14.2 (1.04) & 1 \\ 
 &  &  &  &2 &30.0 & 3.6 & 2.7 &   10.3 &  & 5.2 & 
3.4 (0.75)  & 13.2 (0.75) &  \\ 
G012.7589+00.8954$^M$ & 18:10:06.1 & $-$17:26:34 & 0.07 & 1 &15.6 & 5.6 & 0.6 &    3.6 &  & 6.5 & 
2.1 (1.12)  & 14.5 (1.12) &  \\ 
G012.7720+00.3038$^P$ & 18:12:18.3 & $-$17:42:59 & 0.11 & 1  $^\star$  &18.7 & 3.8 & 3.1 &   12.5 &  & 6.2 & 
2.4 (1.02)  & 14.1 (1.02) & 1 \\ 
 &  &  &  &2 &30.9 & 0.5 & 3.5 &    1.9 &  & 5.2 & 
3.5 (0.74)  & 13.1 (0.74) &  \\ 
 &  &  &  &3 &58.4 & 0.5 & 1.5 &    0.8 &  & 3.8 & 
5.0 (0.43)  & 11.6 (0.43) &  \\ 
G012.7737$-$00.1527$^J$ & 18:13:59.5 & $-$17:56:01 & 0.16 & 1 &45.1 & 1.2 & 16.3 &   20.7 &  & 4.4 & 
4.4 (0.54)  & 12.2 (0.54) &  \\ 
G012.7890$-$00.2185$^J$ & 18:14:15.9 & $-$17:57:06 & 0.27 &1 &10.8 & 1.2 & 2.2 &    2.8 &  & 7.0 & 
1.5 (1.31)  & 15.0 (1.31) &  \\ 
 &  &  &  & 2 &35.5 & 15.5 & 5.6 &   92.0 & Blue shoulder & 4.9 & 
3.8 (0.66)  & 12.8 (0.66) &  \\ 
 &  &  &  &3 &47.8 & 1.2 & 7.8 &    9.9 &  & 4.2 & 
4.5 (0.52)  & 12.1 (0.52) &  \\ 
G012.8062$-$00.1987$^M$ & 18:14:13.6 & $-$17:55:38 & 0.10 &1 &9.1 & 1.5 & 3.8 &    6.0 &  & 7.2 & 
1.3 (1.15)  & 15.3 (1.39) &  \\ 
 &  &  &  & 2  $^\star$  &34.4 & 24.4 & 7.0 &  181.0 &  & 5.0 & 
3.7 (0.68)  & 12.9 (0.68) & 2,3 \\ 
 &  &  &  &3 &47.0 & 2.8 & 9.9 &   29.4 &  & 4.3 & 
4.5 (0.52)  & 12.1 (0.52) &  \\ 
 &  &  &  &4 &59.7 & 0.8 & 2.5 &    2.1 &  & 3.8 & 
5.0 (0.42)  & 11.5 (0.42) &  \\ 
G012.8296$-$00.1719$^P$ & 18:14:10.5 & $-$17:53:37 & 0.17 &1 &11.5 & 0.9 & 6.6 &    6.3 &  & 6.9 & 
1.6 (1.28)  & 14.9 (1.28) &  \\ 
 &  &  &  & 2  $^\star$  &35.4 & 10.4 & 5.8 &   63.9 &  & 4.9 & 
3.8 (0.66)  & 12.8 (0.66) & 3 \\ 
 &  &  &  &3 &48.9 & 3.4 & 10.7 &   38.6 &  & 4.2 & 
4.5 (0.50)  & 12.0 (0.50) &  \\ 
G012.8600$-$00.2737$^P$ & 18:14:36.7 & $-$17:54:57 & 0.16 &1 &12.5 & 2.7 & 4.7 &   13.5 &  & 6.8 & 
1.7 (1.23)  & 14.8 (1.23) &  \\ 
 &  &  &  & 2  $^\star$  &35.8 & 12.0 & 7.0 &   89.0 &  & 4.9 & 
3.8 (0.66)  & 12.8 (0.66) & 1 \\ 
 &  &  &  &3 &45.4 & 2.2 & 2.6 &    6.1 &  & 4.4 & 
4.4 (0.54)  & 12.2 (0.54) &  \\ 
 &  &  &  &4 &51.0 & 3.2 & 4.7 &   15.9 &  & 4.1 & 
4.6 (0.49)  & 11.9 (0.49) &  \\ 
G012.9090$-$00.2607$^M$ & 18:14:39.7 & $-$17:51:59 & 0.46 & 1  $^\star$  &35.8 & 9.7 & 6.7 &   68.9 &  & 4.9 & 
3.8 (0.66)  & 12.8 (0.66) & 2,3,4 \\ 
 &  &  &  &2 &51.8 & 2.8 & 3.9 &   11.6 &  & 4.1 & 
4.7 (0.48)  & 11.9 (0.48) &  \\ 
G013.0105$-$00.1267$^P$ & 18:14:22.2 & $-$17:42:48 & 0.09 &1 &11.8 & 2.7 & 8.1 &   23.2 &  & 6.9 & 
1.6 (1.25)  & 14.9 (1.25) &  \\ 
 &  &  &  &2 &23.3 & 2.8 & 4.6 &   13.7 &  & 5.8 & 
2.8 (0.89)  & 13.7 (0.89) &  \\ 
 &  &  &  &3 &35.5 & 6.9 & 5.8 &   42.4 &  & 4.9 & 
3.8 (0.66)  & 12.8 (0.66) &  \\ 
 &  &  &  &4 &46.2 & 4.1 & 1.6 &    7.0 & Blended & 4.3 & 
4.4 (0.53)  & 12.2 (0.53) &  \\ 
 &  &  &  &5 &50.8 & 4.4 & 10.4 &   48.5 & Blended & 4.1 & 
4.6 (0.49)  & 12.0 (0.49) &  \\ 
G013.1443+05.0644$^J$ & 17:55:45.1 & $-$15:03:42 & 0.13 &\multicolumn{1}{c}{$\cdots$} & \multicolumn{1}{c}{$\cdots$} &\multicolumn{1}{c}{$\cdots$} &\multicolumn{1}{c}{$\cdots$} &\multicolumn{1}{c}{$\cdots$} &\multicolumn{1}{l}{$\cdots$} &\multicolumn{1}{c}{$\cdots$} &\multicolumn{1}{c}{$\cdots$} &\multicolumn{1}{c}{$\cdots$}   \\ 
G013.1840$-$00.1069$^P$ & 18:14:38.8 & $-$17:33:05 & 0.08 &1 &14.6 & 3.2 & 8.0 &   27.1 &  & 6.6 & 
2.0 (1.14)  & 14.6 (1.14) &  \\ 
 &  &  &  &2 &23.9 & 2.4 & 6.4 &   16.3 &  & 5.8 & 
2.9 (0.87)  & 13.7 (0.87) &  \\ 
 &  &  &  &3 &35.9 & 6.3 & 5.2 &   34.7 &  & 4.9 & 
3.8 (0.66)  & 12.8 (0.66) &  \\ 
 &  &  &  &4 &43.4 & 1.8 & 3.2 &    6.1 &  & 4.5 & 
4.2 (0.56)  & 12.4 (0.56) &  \\ 
 &  &  &  & 5 &53.2 & 6.7 & 7.7 &   54.7 & Red asymmetry & 4.1 & 
4.7 (0.47)  & 11.9 (0.47) &  \\ 
G013.2097$-$00.1436$^P$ & 18:14:49.9 & $-$17:32:47 & 0.09 &1 &14.6 & 2.1 & 8.4 &   18.7 & Blended & 6.6 & 
1.9 (1.14)  & 14.6 (1.14) &  \\ 
 &  &  &  &2 &23.3 & 1.6 & 6.1 &   10.3 & Blended & 5.8 & 
2.8 (0.89)  & 13.7 (0.89) &  \\ 
 &  &  &  &3 &36.1 & 2.6 & 6.8 &   18.7 &  & 4.9 & 
3.8 (0.65)  & 12.8 (0.65) &  \\ 
 &  &  &  &4 &44.5 & 2.5 & 2.1 &    5.6 &  & 4.5 & 
4.3 (0.55)  & 12.3 (0.55) &  \\ 
 &  &  &  & 5  $^\star$  &53.1 & 4.6 & 8.8 &   42.9 &  & 4.1 & 
4.7 (0.47)  & 11.9 (0.47) & 1 \\ 

\hline

\multicolumn{14}{l}{$^a$ The superscript letter indicates the telescope used for each observation as follows: (J)CMT, (O)nsala, (M)opra, (F)CRAO and (P)MO.}\\ 
\multicolumn{14}{l}{$^b$ Where multiple components have been detected we identify the most likely component associated with the RMS source by an $\star$.}\\ 
\multicolumn{14}{l}{Notes: (1) strongest component; (2) methanol maser; (3) water maser; (4) and (5) CS detection by \citet{bronfman1996} and \citet{fontani2005}.}\\ 
\end{tabular}
\end{minipage}
\end{sidewaystable*}

\subsection{Identification of multiple components}

\begin{figure}
\begin{center}

\includegraphics[width=0.9\linewidth]{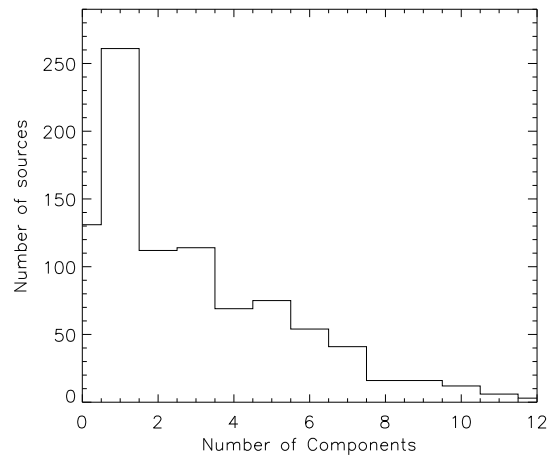}

\caption{Histogram of the number of components detected towards RMS sources reported in this paper.}

\label{fig:comp_hist}
\end{center}
\end{figure}

As previously mentioned two or more components are detected towards the majority of our sources. The number of components detected towards our sources is illustrated in a histogram presented in Fig.~\ref{fig:comp_hist}. Before we can determine the distances to our sources we need to be able to find a way to identify the most likely component associated with each of these MYSO candidates. In some cases the associated component is obvious as it is far stronger than any other component seen in the spectrum, however, identifying the correct component is not always so easy. In the bottom panels of Fig.~\ref{fig:example_spectra} we present two examples of multiple component detections; the left hand panel shows an example of a case where it is easy to identify the component associated with the RMS source, conversely, the panel on the right shows two components that are approximately equal in strength making it difficult to identify the one associated with the RMS source.

One way to solve this component ambiguity is to take advantage of other less abundant molecular line tracers (i.e., CS ($J$=2--1); \citealt{bronfman1996}). CS traces denser gas than $^{13}$CO and should be present in genuine MYSOs rather than more diffuse line of sight molecular cloud material.  However, comparing the single component $^{13}$CO sources detected by us that overlap with sources observed in CS by \citet{bronfman1996} we find that their respective \vlsr's of the rarely disagree by more than one \kms. Searching the \citet{bronfman1996} CS catalogue we were able to resolve the component ambiguity towards 79 RMS sources where an IRAS point sources observed by \citet{bronfman1996} was located within 30\arcsec.

Another method is to use the velocities of water and methanol masers, which are often associated with massive star forming regions, to help identify the component of interest towards some of these multiple component detections. Maser emission can consist of many components spread over a wide range of velocities, typically 10--15~\kms\ for methanol and up to 70 \kms\ for water masers (\citealt{sridharan2002}). Despite these large velocity ranges and number of components, the velocity of the brightest component correlates extremely well with the velocity of the molecular clouds with which they are associated. In Paper~I we compared the \vlsr\ of water and methanol masers from the catalogues of \citet{valdettaro2001} and \citet{pestalozzi2005} with their associated star forming clouds as determined from the CS ($J$=2--1) observations reported by \citet{bronfman1996}. We found a fairly robust correlation, even at rather large angular separations ($\sim$120\arcsec), between the observed velocities of the molecular clouds and the velocities of nearby masers, with a standard deviation of $\sim$2.5~\kms\ and $\sim$3.5~\kms\ for methanol and water masers respectively. It is therefore possible to identify the most likely CO component in cases where the difference in velocity between any two CO components is large and the velocity difference between a particular CO component and the maser velocities is smaller than three times the standard deviation (3$\sigma$ $\simeq$ 7.5 \kms\ and 10.5 \kms\ for methanol and water masers respectively). 

Using a 120\arcsec\  radius we searched  the water and methanol masers catalogues of \citet{valdettaro2001} and \citet{pestalozzi2005} from which we identify 121 methanol masers and 95 water masers associated with 159 RMS sources. Using the 3$\sigma$ distances calculated above as a criterion we are able to determine the most likely component towards 96 RMS sources. Using a combination of archival maser data and CS data we are able to identify the correct component associated with an RMS source in 175 of the 519 multiple component cases. Although using these other tracers of massive star formation has proved useful in reducing the number of multiple component sources by $\sim$40\%, the majority still remain unresolved.

As also found with our southern sources (Paper~I), in nearly every case the maser is associated with the strongest component (i.e., highest integrated intensity (\Trv), and in the few cases where it is not, it is rarely associated with a component less than 50\% that of the strongest component. We have therefore used this additional information to identify the most likely component associated with the MYSO by picking the strongest component as long as it is more than twice as strong as any other component present in the data. However, we note that in a couple of cases the detected maser is associated with a weaker component and it is therefore possible, in a small number of cases, that a component has been incorrectly assigned.   

By applying this criterion to our data we are able to resolve the component ambiguity towards a further 191 sources, bringing the total number resolved to 379 ($\sim$73\%) of the multi-component detections. Together with the single component detections we have been able to determine unique kinematic velocities towards $\sim$80\% of our detections. For sources where we are able to resolve the component multiplicity we indicate the most likely component associated with an asterisk in the component number column of Table~\ref{tbl:source_parameters}. Additional observational data will be required to resolve the remaining 141 sources towards which multi-components are detected. A number of these may be resolved using data from the Methanol Multibeam (MMB) survey\footnote{http://www.jb.man.ac.uk/research/methanol} which is currently underway and plans to survey the whole Galactic plane for $|b| < 2\degr$. For the remaining sources we plan to resolve their component ambiguities with a programme of CS, water masers and CO mapping observations using the GRS data, the results of which will be published in due course.

\subsection{Kinematic distances}
\label{sect:kinematic_distance}

\begin{figure}
\begin{center}
\includegraphics[width=0.9\linewidth]{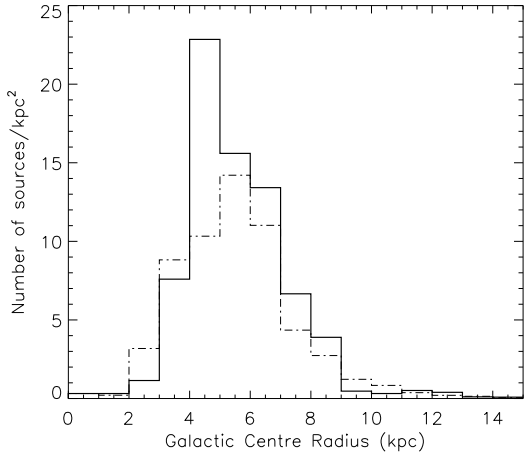}
\caption{Histogram showing the distribution of surface density of clouds with respected to their distance from the Galactic centre. The distribution of northern clouds detected and presented in this paper and southern clouds (i.e., Paper~I) are indicated by the histogram bounded by a solid line and dashed-dotted lines respectively. The bin size for both histograms is 1~kpc. The northern distribution peaks between 4 and 5~kpc consistent with the location of the 5 kpc molecular ring which surrounds the Galactic centre.}

\label{fig:rgc_hist}
\end{center}
\end{figure}

The \citet{brand1993} rotation curve has been used to calculate kinematic distances and Galactocentric radii to all detected components. We have assumed the distance to the Galactic centre to be 8.5~kpc and a solar velocity of 220~\kms. In Fig.~\ref{fig:rgc_hist} we present a histogram of the surface density of all detected components from both the northern (bounded by a solid line) and southern (dashed-dotted line; see Paper~I) Galactic plane as a function of Galactic radii. The distribution of northern components is strongly peaked between 4 and 5~kpc, consistent with the location of the thick ring of material which surrounds the centre of the Galaxy. Comparing the distribution of northern and southern clouds reveals a number of differences; first, the northern distribution peaks slightly nearer to the Galactic centre than that seen in the southern data which peaked between 5 and 6~kpc. The second thing to notice is that the shapes of the distributions are different, with the distribution of the northern clouds being more highly skewed to larger Galactic radii than that of the southern clouds.
  
The rotation curve equation results in two possible solutions for all sources within the solar circle, which includes 90\% of our components. These distances are equally spaced on either side of the tangent point and are commonly referred to as the near and far distances. This is known as the kinematic distance ambiguity problem and makes determining accurate distances to objects located within the solar circle difficult. The only sources within the solar circle that do not suffer from this ambiguity are those located at the tangent points.

The Galactocentric radii and kinematic distances determined from the rotation model for all sources located between 10\degr\ $< l <$ 168\degr\ are presented in columns 11--13 of Table~\ref{tbl:source_parameters}. Our selection criteria excluded sources within 10\degr\ of the Galactic centre bacause of problems of confusion and difficulties in obtaining reliable kinematic distances. We have not presented distances for sources located near the Galactic anti-centre since they are very uncertain (this only applies to 13 sources). 

Sources with components outside the solar circle have only a single distance as do sources located at the tangent points, of which there are 210. Errors in the distances have been determined by shifting the component \vlsr\ by $\pm$10~\kms\ to account for peculiar motions; these result in typical errors of $\sim$1~kpc. If the near and far distances for a particular component are within 1~kpc of each other we place it at the tangent point. In total 203 components were found to be located at or near the tangent point leaving 2171 components with ambiguous distances, however, many of these are not associated with an RMS sources. We intend to resolve the distance ambiguities towards all components associated with MYSO candidates and UCHII regions using HI absorption techniques (e.g., see \citealt{araya2001,araya2002,kolpak2003,jackson2002}) some of which have already been successfully used to resolve distance ambiguities towards a subset of the RMS sample (see \citealt{busfield2006} for details). However, this is beyond the scope of this paper but will be addressed in a subsequent paper (Urquhart et al. in prep.).

\section{Summary}
\label{sect:summary}
We present the results of a programme of molecular line observations of towards 911 candidate massive young stellar objects (MYSOs) located in the 1st and 2nd Quadrants of the Galaxy. The data presented here has been compiled from $^{13}$CO observations made using the JCMT, PMO, Onsala and Mopra telescopes and archival data from the GRS survey. 

Emission is detected towards 780 RMS sources with two or more components being seen towards approximately 56\%. Where available we have used a combination of associated stellar masers (water and methanol) and CS data to identify the cloud associated with an RMS source, and derive a criterion for selecting the most likely cloud associated with each RMS source where no useful tracer is present. In total we have been able to identify the component associated with our MYSO candidates in 379 cases of the 520 multiple detections. Combined with the single component detections we have obtained unambiguous kinematic velocities towards 638 sources ($\sim$80\% of the detections). The 141 sources for which we have not been able to determine the kinematic velocity will require additional line data. No $^{13}$CO emission is detected towards 131 RMS sources; these sources are fairly evenly distributed in both longitude and latitude and since they are not associated with any molecular material are very likely to be evolved stars often confirmed by other characteristics such as their IRAS colours.

These data complement the observations of the southern Galactic plane RMS sources presented in Paper~I. The full data set includes $^{13}$CO observations of 1765 MYSO candidates, and together with archival data provides molecular line data for our entire RMS catalogue. In turn this data set forms part of a larger programme of follow-up observations of near and mid-infrared colour-selected sample of MYSO candidates and are significant step in the delivery of the largest sample of MYSOs to date.

\begin{acknowledgements}

The authors would like to thank the Director and staff at the Paul Wild Observatory, Professor Ji Yang and the staff of the Qinghai station of the Purple Mountain Observatory, as well as the staff of the JCMT (programme ID M03AU35) and Onsala telescope for their hospitality and assistance during our observations. We would also like to thank Thomas Dame for providing the integrated $^{12}$CO maps used in Fig.~1 and Sven van Loo for his comments and suggestions on an early draft of this paper. JSU and CRP are supported by STFC postdoctoral fellowship grants; CRP was also supported by a UNSW Scholarship during the observations. This research would not have been possible without the SIMBAD astronomical database service operated at CDS, Strasbourg, France and the NASA Astrophysics Data System Bibliographic Services. This publication makes use of molecular line data from the Boston University-FCRAO Galactic Ring Survey (GRS). The GRS is a joint project of Boston University and Five College Radio Astronomy Observatory, funded by the National Science Foundation under grants AST-9800334, AST-0098562, \& AST-0100793.

\end{acknowledgements}

\bibliography{scuba2.bib}

\bibliographystyle{aa}

\appendix
\section{Additional $^{13}$CO observations}

In addition to the observations already presented we present images of the fitted spectra and Gaussian parameters obtain towards a further 63 MSX sources. These sources were selected using an early version of the MSX point source catalogue (MSXv1.3), however, when the same slection criteria were applied to a more recent version of the catalogue (MSXv2.3) these sources were no longer found to satisfy our selection criteria and were excluded. Althought these have no direct relevance to the RMS survey they are still likely to be associated with regions of nearby low-mass star formation or more evolved HII regions and we therefore present them here in the hope that they might prove useful to the other members of the community. In Fig.~10 we present images of the observed spectra (black) and the Gaussian fits to the data (red) and in Table~3.\footnote{Fig.~10 and Table~3 are only available in the online version of this journal.}

\end{document}